\documentclass{elsarticle}

\def\benchname{{\tt IMSuite}}
\def\benchnameH{{\bf IMSuite}}
\def\benchexpansion{{\em IIT Madras benchmark suite for simulating distributed algorithms}}
\def\finish{\mbox{\tt finish}}
\def\async{\mbox{\tt async}}
\def\atomic{\mbox{\tt atomic}}

\def\fx10{\mbox{\tt X10-FA}}
\def\cx10{\mbox{\tt X10-FAC}}
\def\fhj{\mbox{\tt HJ-FA}}
\def\phj{\mbox{\tt HJ-FAP}}

\def\MP{\mbox{\tt MP}}
\def\UP{\mbox{\tt UP}}
\def\SP{\mbox{\tt SP}}
\def\maxin{\mbox{\rm Mx-In}}
\def\minin{\mbox{\rm Mn-In}}

\def\afinish{\mbox{\tt Fin}}
\def\abarrier{\mbox{\tt Bar}}
\def\amutex{\mbox{\tt Mut}}

\makeatletter
\newif\if@restonecol
\makeatother

\usepackage{url}
\usepackage{times}
\usepackage{stmaryrd}
\usepackage{colortbl}
\usepackage{graphicx,color}
\usepackage{caption}
\usepackage[labelformat=simple]{subcaption}
\usepackage[linesnumbered]{algorithm2e}
\usepackage{float}
\usepackage{natbib}

\DeclareCaptionType{copyrightbox} 
\usepackage{array}

\newcommand{\ifTR}[1]{} 
\newcommand{\ifConf}[1]{#1} 
\usepackage{code}

\begin{document}

\title{IMSuite: A Benchmark Suite for Simulating \\Distributed Algorithms}
\author{Suyash Gupta}
\ead{suyash@cse.iitm.ac.in}

\author{V. Krishna Nandivada\corref{nvk}}
\ead{nvk@cse.iitm.ac.in} 
\cortext[nvk]{Corresponding author}

\address{PACE Lab, Department of Computer Science and Engineering, IIT Madras, India 600036\\phone/fax: +91-44-22574380\clearpage}

\newpage

\begin{frontmatter}
\clearpage
\begin{abstract}
\label{s:abstract}
%
Considering the diverse nature of real-world distributed applications that 
makes it hard to identify a representative subset of distributed benchmarks,
we focus on their underlying distributed algorithms.
We present and characterize a new kernel benchmark suite (named \benchname{}) 
that simulates some of the classical distributed algorithms in task
parallel languages.
We present multiple variations of our kernels, broadly categorized under two heads:
(a) varying synchronization primitives (with and without fine grain synchronization
primitives); and
(b) varying forms of parallelization (data parallel and recursive task parallel).
Our characterization covers interesting aspects of distributed
applications  such as 
distribution of remote communication requests,
number of synchronization,
task creation,
task termination and atomic operations.
We study the behavior (execution time) of our kernels by
varying the problem size, the number of compute threads, and the input configurations.
We also present an involved set of
input generators and output validators.

\end{abstract}


\end{frontmatter}
\newpage
\section{Introduction}
\label{s:intro}

Large distributed applications find their use in a variety of diverse domains:
banking, telecommunication, scientific computing, 
network on chips, and so on.
The diverse and complex nature of these distributed applications makes it
hard to identify a representative subset of distributed benchmarks.
The absence of such a benchmark set hinders the design of new
optimizations and program analysis techniques that can be applied
uniformly across many distributed applications.

The common denominators of most of the distributed applications are the
underlying distributed algorithms.
Both the distributed applications and the underlying distributed algorithms
display common traits such as communication, timing and failure.
We argue that compared to the complex distributed applications, reasoning
about these underlying algorithms can be easier and can also help
in analyzing the diverse applications that use them.
Thus, we believe that a kernel benchmark suite implementing popular distributed
algorithms is in order.

We now lay down a set of seven {\em key} requirements necessary for such a kernel
benchmark suite.
These requirements are categorized under the following three heads.\\
\noindent
(A) {\em Requirements based on characteristics of kernel benchmarks
implementing distributed algorithms}:
Our study of popular text books~\citep{lynch,PelegBook} and lecture
notes~\citep{roger} on distributed algorithms helped us derive the important
characteristics of typical distributed algorithms;
these characteristics form the basis of our first three key requirements.

\begin{enumerate} 
\item The algorithms implemented by the kernel benchmarks must solve 
common challenges in distributed systems.

\item The kernels should cover important characteristics of 
distributed systems such as communication (broadcast, unicast, or
multicast), timing (synchronous, asynchronous or partially
synchronous) and failure.

\item The benchmark kernels should simulate the behavior of distributed
systems consisting of (partially) independent nodes and the interconnect
thereof.

\end{enumerate}

\noindent
(B) {\em Requirements based on the target hardware}: 
\begin{enumerate}
\setcounter{enumi}{3}
\item The execution of the kernels implementing distributed algorithms should not necessarily require a
complex hardware setup; these should be usable in the presence of a
shared memory multicore / distributed memory multicore or even a sequential system.

\end{enumerate}

\noindent
(C) {\em Requirements based on best practices in existing benchmark suites}:
The final two requirements are derived from the best practices followed in
well known benchmark suites, such as
PBBS~\citep{pbbs}, NPB~\citep{npb},
BOTS~\citep{bots}, PARSEC~\citep{parsec}. 

\begin{enumerate}
\setcounter{enumi}{4}
\item  The kernels should be small in size and easy to debug.
\item The benchmark suite should provide a variety of inputs (with varying
configurations and sizes) and convenient means to verify the generated
output.
\item The benchmark suite should provide means to analyze static and dynamic
characteristics specific to the domain under consideration (distributed
systems in our case).
\end{enumerate}

%
%
%
%
%
%
%

%

Our study of existing benchmark suites%
~\citep{
pbbs,npb,bots,parsec,epccmpi,berl,specomp,epcc,jgf,nasgb,intelmpi,hpcc,specmpi,splash92,splash
}~has
found that none of them meet majority of the aforementioned key requirements.
Our goal is to design a benchmark suite that meets all our
stated requirements. 
As a first step, we shortlist a set of important problems 
in the context of distributed systems.

\begin{figure}[t]
\small
\begin{tabular}{|p{0.21\columnwidth}| p{0.70\columnwidth}|} \hline
 Problem 		&	Applications and/or domains of interest.\\ \hline
                         	
 Breadth~first~search		& 	semantic graphs and 
					community analysis.\\ \hline                        
 Consensus 			& 	checking reliability of  a system; 
 					grid computing; peer-to-peer
					networks; sensor networks. \\ \hline
                         	
 Routing table 			& 	internet routing tables; OSPF protocol.\\ \hline
                         
 Dominating set 		&	mobile adhoc network routing;
 mobile wireless adhoc networks. \\ \hline
 
 Maximal independent set	& 	symmetry breaking in networks;
					clustering in wireless ad hoc and
					sensor networks.  \\ \hline
                         
 Committee~creation 		& 	dynamic networks; 
 					token dissemination protocol.\\ \hline
                         
 Leader election 		& 	selecting a coordinating node for a network.\\ \hline
                         
 Spanning tree 			& 	in IEEE 1394.1 standard for 
 					interconnecting LANs using
					bridges.\\ \hline
                         
 Graph coloring 		& 	color growth bounded graphs such
 					as unit disk graphs; 
					for resolving resource conflicts. \\ \hline

\end{tabular}
\caption{\small Core distributed computing problems and their applications.}
\label{fig:algos-apps}
\end{figure}
Figure~\ref{fig:algos-apps} shows some of the {\em core}
problems in the area of distributed computing and lists
a few of their diverse applications.
The centrality of these problems can also be seen from the importance
given to them in popular 
textbooks and lecture notes on distributed
algorithms~\citep{lynch,PelegBook,roger}.
In this paper, we present and characterize a new kernel benchmark suite 
named \benchname{}: \benchexpansion{}
that implements some of the classical algorithms 
to solve these core problems;
we refer to these algorithms as the {\em core} algorithms.

\benchname{} implements the core algorithms in two task parallel languages
X10~\citep{x10manual} and HJ~\citep{hjava}.
X10 and HJ languages with their APGAS-model to easily simulate the
distributed systems, light weight tasks to represent the computation in
the distributed nodes, and clocks/phasers to model lock step synchrony in
irregular and recursive applications, give a convenient way to program
distributed kernels.  
One of the main advantages of using these languages is that they can easily 
simulate a large set of distributed nodes even in the absence of 
complex distributed hardware.  

\noindent
{\bf Our contributions}\\
\noindent
$\bullet$
We present a study of a large set of existing benchmark suites and discuss
their limitations with respect to our stated key requirements
(Section~\ref{s:related}).\\
\noindent
$\bullet$
We give a methodical approach to implement distributed algorithms in 
task parallel languages to run on hybrid systems%
\footnote{A hybrid system may consist of one or more distributed nodes
(with a capability to communicate with each other),
each node may consist of one or more cores, and each core in turn may have one or
more hardware threads.
} (Section~\ref{s:trans}).\\
\noindent
$\bullet$ 
Considering the different popular parallel programming styles, we
present multiple variations of our kernels in both X10 and HJ.
These variations (totaling 31 per language) can be broadly
categorized under two heads:
(a) {\em Varying synchronization primitives}: 
Our benchmark kernels can use fine grain synchronization primitives 
(such as phasers in HJ and clocks in X10), or can realize
synchronization by joining/terminating each task and recreating them
later. 
(b) {\em Varying forms of parallelization}:  \benchname{} contains a data parallel implementation
for each core algorithm.
Further, \benchname{} also 
includes recursive task parallel versions for some of the core algorithms.
Besides these parallel versions \benchname{} also includes the
corresponding serial
implementations (Section~\ref{s:implement}).
\\
%
\noindent
$\bullet$
We provide an algorithm specific input generator that can
generate a variety of inputs with varying configurations.
Each benchmark also includes an output validator.\\
\noindent
$\bullet$
We characterize  \benchname{}  on a hybrid system. 
Our characterization covers interesting aspects of distributed
applications  such as 
distribution of remote communication requests,
number of synchronization,
task creation,
task termination and atomic operations.
We study the behavior (execution time) of our kernels by
varying the problem size, the number of compute threads, and the input configurations.
(Section~\ref{s:eval}).

%
%


%




\section{Related Work}
\label{s:related}
In this section we categorize some of the
popular benchmark suites catering to parallel and distributed systems and
discuss their limitations with respect to our stated key requirements.

\noindent{\bf Applications vs. kernels vs. micro-kernels}  -
Many of the well-known benchmarks consist of a set of representative
applications.
Examples include 
NPB~\citep{npb},
BOTS~\citep{bots},
PARSEC~\citep{parsec} (including its two prior {\em avatars}  SPLASH~\citep{splash92} and SPLASH-2~\citep{splash}),
BenchERL~\citep{berl}, 
SPEC-OMP~\citep{specomp},
JGF~\citep{jgf},
NGB~\citep{nasgb},
SPEC-MPI~\citep{specmpi}
and 
HPCC~\citep{hpcc}. 
While PARSEC, JGF, NPB and BenchERL also contain a few kernels,
benchmark suites like PBBS~\citep{pbbs} focus only on kernel benchmarks.
Similarly, JGF contains a few micro-kernels as well, while 
EPCC~\citep{epcc,epccmpi} and IntelMPI~\citep{intelmpi} contain only micro kernels.
Compared to the application-oriented benchmarks, the kernel benchmarks 
are small in size, simpler to understand, easier to debug
and provide insights on how a certain algorithm behaves.
Micro-kernels on the other hand are helpful to study 
a specific feature of a language, runtime or architecture.\\
%
\noindent{\bf Scientific vs. non-scientific} -
Most of the parallel benchmark suites 
target scientific or mathematical computations.
Examples include BOTS, JGF, HPCC, NPB, SPEC-OMP, SPEC-MPI and PARSEC.
The PBBS benchmark suite consists of a mixed bag of scientific and graph 
computations.
Benchmark suites like BenchERL, EPCC, and IntelMPI consist of mainly synthetic benchmarks. 
\\%
\noindent{\bf Task parallel vs. loop parallel vs. recursive} -
BOTS and PBBS admit both task parallel and recursive task parallel
computation. 
In contrast, JGF, HPCC, SPEC-OMP, SPEC-MPI, IntelMPI, PARSEC, NPB, NGB,  and EPCC
suites include computations chiefly depicting loop level parallelism.
\\%
\noindent{\bf Parallel vs. hybrid systems} -
Benchmark suites like NPB, JGF, NGB, BenchERL, SPEC-MPI, EPCC, and IntelMPI contain benchmarks
that can run over hybrid systems, while rest of the benchmark suites can run only on a parallel system.

Of the seven key requirements discussed earlier in the section, the first
four are specific to the distributed applications and hence are not
satisfied by any of the discussed benchmark suites.
The PBBS benchmark suite satisfies Req\#5 and Req\#6.
On the other hand,
PARSEC, JGF, NPB and BenchERL have a mix of small kernels and large
applications, and satisfy Req\#5 partially.
Similarly,
benchmark suites such as BOTS, BenchERL, HPCC and SPEC include an output 
verifier for a pre-defined input, and satisfy Req\#6 partially.
Req\#7 is partially satisfied by
PBBS, PARSEC, SPLASH and BOTS -- they allow the user to measure some
dynamic characteristics that are pertinent to parallel programs (such as,
number of tasks, barriers, joins, and so on).

Compared to these benchmark suites, \benchname{} satisfies all the key requirements.
It consists of kernels that implement 
popular distributed algorithms (mostly graph based) that are mainly
irregular and non-scientific in nature.
The kernels in \benchname{} exhibit both loop and recursive task parallelism.
While the current implementation of \benchname{} is in X10 and HJ,
these kernels can easily be ported to other languages that support
appropriate runtime models (such as APGAS or global address space).

\section{Background}
\label{s:back}
\subsection{Core algorithms}
\label{s:bench}
\begin{figure*}
\small
{
\begin{center}
\begin{tabular}{|c|c|c|c|} \hline

 Problem Category 	(Abbr)		& NW type 	& Time Complexity 			& Message Complexity 						\\ \hline	
                                                                                                                                                                
 Breadth First Search	({\em BF}) 	& General 	& $O(D)$ 				& $O(nm)$ 							\\ \hline
 Breadth First Search	({\em DST})	& General 	& $O(D^2)$ 				& $O(m+nD)$ 							\\ \hline
 Consensus(Byzantine) 	({\em BY}) 	& General 	& $O(D)$                 		& $O(n^2)$ 							\\ \hline
 Routing Table Creation ({\em DR})	& General 	& $O(n^2)$	 			& $O(nm)$ 							\\ \hline
 Dominating Set 	({\em DS}) 	& General 	& $O(\log^2 \Delta$ 			& $O(n\times \Delta ^2$						\\
			      	&		& $\times \log n)$ 			& $\times \log^2\Delta \times \log n)$ 				\\ \hline
 Maximal Independent Set ({\em MIS})	& General 	& $O(\log n)$ 				& $O(m\log n)$ 							\\ \hline
 Committee Creation 	({\em KC}) 	& General 	& $O(K^2)$ 				& $O(K^2 m)$ 							\\ \hline
 Leader election 	({\em DP}) 	& General 	& $O(D)$ 				& $O(Dm)$ 							\\ \hline
 Leader election 	({\em HS}) 	& Ring (bi) 	& $O(n)$ 				& $O(n\log n)$ 							\\ \hline
 Leader election 	({\em LCR})	& Ring (uni)	& $O(n)$ 				& $O(n^2)$ 							\\ \hline
 Spanning Tree 		({\em MST})	& General 	& $O(n \log n)$ 			& $O(m \log n)$ 						\\ \hline
 Vertex Coloring 	({\em VC}) 	& Tree 		& $O(\log^{*}n)$ 			& $O(n \log^{*}n)$ 						\\ \hline

\end{tabular}
\end{center}
\caption{Core algorithms and their characteristics. Notation: $n$ denotes the number nodes, $m$ denotes the number of edges, and $D$ denotes the diameter, $K$ denotes the maximum committee size and $\Delta$ denotes the maximum degree of the graph.}

%

\label{fig:bench-characteristics}
}
\end{figure*}

We now briefly describe some of the popular algorithms that solve
the problems discussed in Figure~\ref{fig:algos-apps}.
Figure~\ref{fig:bench-characteristics} presents some characteristics of these algorithms.

\noindent
{\bf Breadth First Search (BFS)}:
\ifConf{
We use two different BFS algorithms {\em
BF}~\citep{roger} and {\em DST}~\citep{roger}.
While {\em BF} outputs the distance of every node from the root,
{\em DST} outputs the BFS tree.
\\
}
\ifTR{
We use two different breadth first search algorithms. 
\begin{enumerate}
\item 
{\bf 1. Bellman and Ford (BF)} 
This algorithm~\citep{roger} finds the breadth first spanning tree 
in a general network for a given root node; 
it outputs the distance of every node from the root.
\ifTR{
The algorithm starts by marking the root node and setting its {\em distance value} 
($dValue$ in short) to 0.
The root node  then sends ($dValue + 1$) it to its neighbors.
If a node receives the $dValue$ for the first time or a smaller $dValue$ than what it 
had before then it updates its $dValue$.
Each such node sends ($dValue + 1$) to their neighbors.
The algorithm terminates after each node receives a minimum $dValue$ possible 
and thus it does not further propagate its $dValue$.
}

\item 
{\bf 2. Dijkstra (DST)}  
This algorithm~\citep{roger} finds the 
breadth first spanning tree in a general network;
it outputs the BFS tree that marks the parent and child (if any) node for each node
in the network.
Compared to BF, DST has a higher time complexity and lower message complexity.
\ifTR{
The algorithm starts by marking the root as the {\em current tree}.
In each phases $p$, nodes at a distance $p$ from the root are added to the current tree.
Every newly added node {\em echoes} its arrival by sending an acknowledgment 
(an ACK signal) to its parent.
Each node receiving an ACK signal, echoes the ACK signal to its parent.
The algorithm terminates as soon as no new node is found by the root.
}
\end{enumerate}

}
\noindent
{\bf Byzantine Agreement}:
The byzantine agreement ({\em BY}) algorithm~\citep{motwani} builds a consensus
among the ``good'' nodes of a network that may also contain
``faulty'' nodes. 
\ifTR{
In each iteration, the good nodes share their vote to their neighbors, the fault ones may send
any arbitrary vote.
Each node sets its own vote for the next iteration
based on the votes it received from its neighbors in the current iteration.
Say there are a set of $n$ nodes (or generals) who wish to reach a consensus. 
But there is a set of $t$ faulty nodes (i.e. rest $n-t$ nodes are good nodes)  which 
affect the system.
The $t$ faulty nodes can send different messages to different neighbors.
The ``good'' nodes on the other hand always send the same decision to all their neighbors.
The algorithm guarantees that all the non-faulty nodes reach a consensus at the end.
}
\\
\noindent{\bf Routing}:
In the Dijkstra routing ({\em DR}) algorithm~\citep{tanenbaum} 
each node in the network works independently and 
computes a routing table in parallel. 
\ifTR{
The routing table contains information such as total cost of the path, 
next node on the path to the destination, total hop counts, and so on.
}
\\
%
\noindent{\bf Dominating Set}:
The dominating set ({\em DS}) algorithm~\citep{roger} creates a dominating
set using a probabilistic method that depends on the first and second level 
neighbors of a node.
\ifTR{
The dominating set algorithm~\citep{roger} creates a dominating set by coloring the nodes 
in a specific way:
initially color of each node is assumed to be {\em White} (= not part of a dominating set).
As a node becomes part of the dominating set then its color changes to {\em Black} while 
the color of all its neighbors is set to {\em Grey}. 
A node $n$ is probabilistically added to a dominating set  if its first and
second level neighbors do not have more {\em White} neighbors than $n$.
}
\\
\noindent{\bf {\large $k$}-Committee}:
For a given integer value of $k$, the $k$-committee ({\em KC})
algorithm~\citep{roger}
partitions the input nodes 
into committees of size at most $k$.
\ifTR{
Initially all nodes have a key value. 
The algorithm executes for $k$ phases.
In the first phase, some of the nodes are designated as leaders -- the nodes whose
key value is the minimum compared to all its neighbors at a distance~$\leq k$.
In each subsequent phase, every node which is not associated with any
committee
and whose key value is the minimum compared to all its non-leader neighbors at
a distance~$\leq k$ associates itself with a committee.
Say the size of the largest committee in phase $i$ is $S_i$
and the size of the largest committee in phase $i+1$ is $S_{i+1}$.
Then the algorithm guarantees that $S_{i+1} \leq S_i + 1$.
}
\\
\noindent{\bf Maximal Independent Set}:
{\em MIS}~\citep{roger} uses a randomized algorithm to compute the maximal
independent set for a given input graph.
\ifTR{
In each iteration, every node selects a random value and communicates the same
to all its neighbors.
A node $n$ adds itself to the MIS if its random value is the smallest one among all of its
neighbors.
Thereafter, all the neighbors of $n$ do not contend for the membership of MIS.
}
\\
\noindent{\bf Leader Election}:
We consider three different leader election algorithms.
\ifConf{
The {\em LCR} and {\em HS} algorithms~\citep{lynch} work on a set of nodes organized in
a ring network, where the data flow is unidirectional and bidirectional,
respectively. 
Compared to that, the {\em DP} algorithm~\citep{Peleg} 
works on a set of nodes organized in any general network. 
}
\ifTR{
In each of these algorithms,
every node in the network has
an unique identifier and at the end of the algorithm the node with the highest 
identifier is selected as the leader.

{\bf 1. LCR }
This algorithm~\citep{lynch} finds the leader for a set of nodes organized in
a ring network, where the data flow is unidirectional. 

{\bf 2. HS } 
This algorithm~\citep{lynch} finds the
leader from a set of nodes organized in a ring network, where the data flow is bidirectional. 

{\bf 3. DP }
This algorithm~\citep{Peleg} 
finds the leader for a set of nodes organized in any general network. 
We have implemented a synchronous version of the proposed algorithm. 
The algorithm computes the leader and the diameter of the network.
}
\\
\noindent{\bf Minimum Spanning Tree}:
The {\em MST} algorithm~\citep{roger} works on a
weighted graph.
It starts by marking  every node as an independent fragment,
and proceeds by joining fragments along the {\em minimum
weighted edge}, till a lone fragment is left.
\\
\noindent{\bf Vertex Coloring}: The vertex coloring ({\em VC}) algorithm~\citep{roger}
 colors the
nodes of a tree with three colors.  It first colors the tree using six colors
using a fast algorithm $O(\log^* n)$ and then uses a {\em shift down}
operation (constant time) to color the tree using three colors.


\subsection{X10 and HJ background}

Figure~\ref{fig:x10-cheat-sheet} presents some constructs 
of X10 relevant to this manuscript
(see the language manual~\citep{x10manual} for details).


\begin{figure}[t]
	\small
\begin{tabular}{p{0.34\columnwidth}|p{0.56\columnwidth}}
Syntax & Explanation\\\hline
{\tt{[clocked] async \hspace*{2cm} \{S1\}}} & 
 spawns a new asynchronous task to execute the statement {\tt S1}.
 The {\tt clocked} option registers the task on the set of clocks held by the current task.
\\\hline
{\tt{[clocked]~finish \hspace*{2cm} \{S1\}}}&
 waits for all the tasks created within {\tt S1} to terminate. 
 The {\tt clocked} option introduces a new clock
 that the task executing {\tt S1} gets registered to.\\\hline
{\tt atomic \{S1\}}&
 updates the shared data in {\tt S1} in an atomic fashion, provided 
 other possible accesses to that shared data also happens inside an {\tt atomic}.\\\hline
{\tt Clock.advanceAll}&
 a blocking call that advances all the clocks the current task is
 registered with. It acts as a barrier.\\\hline
{\tt x = at(p) \{S1\} } &  
executes the expression {\tt S1} at place {\tt p} and {\tt x} stores
the return value. \\\hline
{\tt var R: Region; R~=~0..(n-1)} & 
 creates a region $R$ containing $n$ elements: {\tt 0\ldots{}n-1}.\\\hline
\end{tabular}
\caption{X10 command cheat sheet.}
\label{fig:x10-cheat-sheet}
\end{figure}

We use {\tt async} to spawn a new task, {\tt finish} to join tasks, and
{\tt atomic} to provide mutual exclusion.
X10 provides an abstraction of a {\tt Clock} that helps tasks make
progress in lock step synchrony.
A task may be registered on one or more clocks and all the tasks registered on a clock 
make progress in lock step by {\em advancing} the clocks.
A clock is considered to have advanced to the next ``clock tick" if all the
tasks registered on that clock have requested the advancement of the clock.

In X10, a {\tt place} abstracts the notion of computation (multiple tasks)
and data (local to the place).
The set of places available to a program are fixed at the time the program
is launched.
The {\tt at} construct can be used to access remote data.

A {\em region}  is used to represent the
iteration space of loops and the domain of arrays.
A {\em distribution} maps the elements of a region to the set of runtime places.


%
\noindent


\noindent
{\bf Comparison with HJ}:
The parallelism related constructs of 
Habanero Java~\citep{hjava} are similar to that of X10 with minor differences in
syntax and semantics.
For example, HJ constructs {\tt isolated}, {\tt next}, and {\tt phaser} map to 
corresponding X10 constructs {\tt atomic}, {\tt Clock.advanceAll} and {\tt
clock} (refer to the HJ manual~\citep{haban} for details).

\section{Transformation Scheme}
\label{s:trans}
\label{salienttrans}
We now present an overview of our scheme for implementing
distributed algorithms in a task parallel language, to be executed on a
hybrid system.
We list some of the main abstractions pertinent to distributed
algorithms and lay down a procedure for their implementation.\\
\noindent{\bf Node}:
A {\em node} in a distributed algorithm requires 
some data (such as a unique identifier, a mailbox,
information about neighbors and so on) and performs some computations.

The computation of a node can be abstracted by one or more parallel
tasks, in task parallel languages. 
A distributed node (including both computation and data) can be abstracted
by an X10 {\em place} running only the task(s) corresponding to
that node;
we refer to it as the {\em Unique-Place~(\UP{}) model}.
However, such an abstraction can pose a challenge in languages such as
Java and OpenMP that neither support a notion like {\em places} nor can run on
distributed memory systems.
Programs in these languages can be seen as running all the tasks
(corresponding to all the nodes) at a single place.
This simulates a particular type of distributed system where all the data
is ``local'' and the inter-node communication cost is minimal.
X10 and HJ can mimic this behavior when they are
restricted to run on a single place; we term it as the {\em
Single-Place~(\SP{}) model}.
We can also consider a more general scenario, where the runtime consists
of multiple places and each place may simulate the tasks corresponding to
more than one node; we term it as the {\em Multi-Place~(\MP{}) model}.
\\
\noindent
{\bf Communication}:
A set of distributed nodes communicate with each other through message transfer.
These messages are transferred, from one node to another, along the links 
of the underlying network.

Our simulation of the transfer of data between two connected nodes of a network
depends on the runtime model (\MP{}, \UP{} or \SP{} model).
In the context of an \SP{} model, the data transfer is done using the 
shared memory. 
On the other hand,
in the context of \MP{} and \UP{} model, the data transfer may involve message passing.
%
%
%
\\
%
\noindent{\bf Timing}:
In a distributed system, nodes can work asynchronously or synchronously. 
In a synchronous setting there is an assumption of existence of a global 
clock and the nodes proceed in a lock-step fashion, synchronized over 
the global clock.
Contrast to that, in an asynchronous setting there is no concept of a global 
clock and each node works independently.

We achieve lock step synchrony by using fine grain synchronization primitives 
(such as phasers in HJ and clocks in X10) or by 
repeated task-join and task-creation operations (See
Section~\ref{ss:sample-trans} for an example).
\\
\noindent{\bf Phases and Rounds}:
Distributed algorithms are organized around the notion of phases and 
rounds; 
each phase consists of one or more rounds. 
Phases and rounds can be implemented using serial loops.
\\
\noindent{\bf Messages and Mailbox}:
Nodes in a distributed system communicate by exchanging messages.
The size and structure of a message depend on the underlying
algorithm.
Each message is delivered to the receiver's mailbox (a FIFO queue).
The design of the mailbox must ensure that it can hold all the
messages required at any point of time.


\subsection{Sample transformation}
\label{ss:sample-trans}
\begin{figure}[t]
	\small
\SetVline
\restylealgo{boxed}  
\linesnotnumbered
\begin{algorithm}[H]
  {\bf Input} {$n$ nodes each having a unique $uid$};
  {\bf Output} Leader
 
  \For{$round \leftarrow 1$ \KwTo $n$}{ 
  \For{$j\leftarrow 1$ \KwTo $n$ nodes {\bf\rm in parallel}}{
	transmit {\em send}$_j$ to its clockwise neighbor \;
		
		$x \leftarrow $ incoming message\;
		\lIf{$x$ $>$ {uid$_j$}} {{\em send}$_j$ $\leftarrow$ $x$} \;
		\lElseIf{$x$ = uid$_j$} {{\em status$_j$} $\leftarrow$ leader}\;
		\lElse{do nothing} \; }	}
\end{algorithm}
\caption{Distributed Leader Election {\em LCR} Algorithm.}
\label{alg:dist-leader-elect}
\end{figure}
In this section we illustrate our transformation scheme using an example.
Figure~\ref{alg:dist-leader-elect} presents the core of the {\em LCR} algorithm
(see Section~\ref{s:bench}).
Each node $j$ contains three fields: 
{\em uid$_j$} (the unique identifier of the node), {\em send$_j$}
(identifier of the leader as per node $j$ -- initialized to {\em uid}$_j$),
and {\em status$_j$} (if it is a leader or a common node).
The {\em LCR} algorithm runs for $n$ rounds. In each round 
every node $j$ sends {\em send}$_j$ to its neighbor
(successor) and receives the incoming message from its predecessor.
Since {\em LCR} algorithm works on a uni-directional ring network, a node can 
at most receive one message per round.
This allows us to set the size of the mailbox of each node to one.


We now briefly explain how we derive a benchmark kernel for the {\em LCR} algorithm.
\ifConf{
For illustration purposes we use \cx10{} as the target language and later 
state the differences between a code written in \cx10{} and in \fx10{}.
}
\ifTR{
For illustration purposes we use \fx10{} and \cx10{} as the target languages.
}
%
%
The structure of the {\em Node} class for our implementation of the {\em LCR}
algorithm is shown in Figure~\ref{fig:lcr-class}.
We note two interesting points: a) the
{\tt mbox} field can hold at most one message, b) the {\tt nextIndex} field can be eliminated by
some smart design decisions (e.g. in a unidirectional ring network the {\tt nextIndex} of the $j^{th}$ node
can be set to $j+1$).


%
\begin{figure}[t]
\small
\begin{tightcode}
 class Node \{
       var uid:Int;
       var mbox:Int;         // mail box
       var nextIndex:int;    // neighbor index
       var status:boolean;   // true => leader
       var send:Int;         // outgoing message
       var leader:Int; \}
\end{tightcode}
\caption{Structure of the abstract node for LCR} 
\label{fig:lcr-class}
\end{figure}
\begin{figure}[H]
\small
\begin{tightcode}
leader_elect_lcr(val n:Int)\{
   var R:{\bf Region} = 0..(n-1) ;
   var D:{\bf Dist} = {\bf Dist.makeBlock}(R);
   var ndSet:{\bf DistArray} = {\bf{DistArray.make}}[Node](D);
   for(var round:Int=0; round<n; round++) \{
      {\bf clocked finish} \{
         for(j in D)\{ // run each node
            {\bf clocked async at}(D(j))\{ // in parallel
                Transmit(ndSet(j).send, nextIndex);
                {\bf{Clock.advanceAll}}() ;
                if(ndSet(j).mbox > ndSet(j).leader) \{
                   ndSet(j).send = ndSet(j).mbox;
                   ndSet(j).leader = ndSet(j).mbox; \}
                elseif(ndSet(j).mbox==ndSet(j).uid) \{
                   ndSet(j).status = true;
                   ndSet(j).leader = ndSet(j).uid;
                \} \} \} \} \} \}

Trasmit(val receiver:Point, val msg:Int)\{
   {\bf at}(D(receiver)) ndSet(receiver).mbox=msg;\}
                    (a)

leader_elect_lcr(val n:Int)\{
   \ldots
   for(var round:Int=0; round<n; round++)\{
      {\bf finish}\{
         for(j in D)\{ // run each node
            {\bf async at}(D(j))\{ // in parallel
                Transmit(ndSet(j).send, nextIndex);
             \} \} \}
      {\bf finish}\{
         for(j in D)\{ // run each node
            {\bf async at}(D(j))\{ // in parallel
                if(ndSet(j).mbox > ndSet(j).leader)\{
                  \ldots \}
                elseif(ndSet(j).mbox==ndSet(j).uid)\{
                  \ldots
              \} \} \} \} \}
                    (b)
\end{tightcode}
\caption{(a) Core of the LCR algorithm in \cx10{}; 
	 (b) core of the LCR algorithm in \fx10{}; only differences with respect to the \cx10{} version are shown. 
	 The X10 specific constructs are shown in {\bf bold}. See Section~\ref{s:back} for X10 syntax.}
\label{fig:lcr-cx10}
\end{figure}

Figure~\ref{fig:lcr-cx10}(a) shows the core of the {\em LCR} algorithm in \cx10{}.
It creates a blocking distribution {\tt D} over a
region {\tt R} (of $n$ points, where $n$ = number of nodes) and allocates
the array {\tt ndSet} (of $n$ elements), distributed over {\tt D}.
The number of blocks in the distribution {\tt D} is 
set to the number of {\em places} at runtime. 
In each round, the parallel task corresponding to each node transmits
its message and waits (by using  {\tt Clock.advanceAll}) for the
message from its neighbor.
After that, each task recomputes the leader related information based on
the received message and proceeds to the next round.

Figure~\ref{fig:lcr-cx10}(b) sketches the \fx10{} kernel for {\em LCR},
showing only the differences with respect to the \cx10{} kernel shown
in Figure~\ref{fig:lcr-cx10}(a).
This \fx10{} version uses repeated task-join and task-creation operations 
to synchronize the tasks corresponding to the nodes. 
\cx10{} implementation can be considered lightweight as it uses fewer number 
of task creation ({\em async}) and join ({\em finish}) operations and 
utilizes the lightweight synchronization operations offered by {\em Clocks}.

Compared to {\em LCR}, where the \fx10{} and \cx10{} implementation are not 
much different, there are other kernels (such as {\em DS}) where the
differences are significant.
This is especially true when the clock based synchronization operations
are nested deep inside conditional or looping constructs.




\ifTR{
\begin{figure}
\begin{tightcode}
leader_elect_lcr(val n:Int)\{
 var R:{\bf Region} = 0..(n-1) ;
 var D:{\bf Dist} = {\bf Dist.makeBlock}(R);
 var ndSet:{\bf DistArray}
	={\bf DistArray.make}[Node](D);
 for(var round:Int=0; round<n; round++)\{
  {\bf clocked finish}\{
   for(j in D)\{
    {\bf clocked async at}(D(j))\{
      Transmit(ndSet(j).send, nextIndex);
     \} \} \} 
  {\bf clocked finish}\{
   for(j in D)\{
    {\bf clocked async at}(D(j))\{
      if(ndSet(j).mbox > ndSet(j).leader)\{
       ndSet(j).send = ndSet(j).mbox;
       ndSet(j).leader = ndSet(j).mbox;
      \}
      elseif(ndSet(j).mbox==ndSet(j).uid)\{
       ndSet(j).status = true;
       ndSet(j).leader = ndSet(j).uid;
      \} \} \} \} \} \}
Trasmit(val receiver:Point, val msg:Int)\{
 {\bf at}(D(receiver)) 
   ndSet(receiver).mbox = msg;\}       	
\end{tightcode}
\caption{Implementation of LCR in \fx10{}}
\label{fig:lcr-fx10}
\end{figure}

Figure~\ref{fig:lcr-fx10} shows the core of the LCR algorithm in \fx10{}.
\ldots	
}

 
\section{Internals of \benchnameH{}}
\label{s:implement}
In this section, we briefly explain the internal details of \benchname{}. 
This benchmark suite implements the twelve core algorithms 
described in Section~\ref{s:bench}.
Considering the different popular parallel programming styles, we
have implemented multiple variations of these algorithms:

{\em Varying the synchronization primitives}: 
We implement all our variations in two subsets of X10: 
(a) \fx10{} -- uses the \finish{}, \async{} and \atomic{} constructs for task
creation, join and mutual exclusion, respectively.  Synchronization is achieved
by joining/terminating all the tasks and recreating them later.
(b) \cx10{}  -- uses the abstraction of {\em clocks} in addition to the
constructs of \fx10{}.
Clocks provide efficient synchronization primitives that can be used to yield
arguably more compact and efficient programs.
All the core algorithms (except {\em DR}) have been implemented in both \fx10{} and \cx10{}.
In case of {\em DR},
we found no scope of using low level synchronization primitives like clocks.
And hence we have this algorithm implemented only in \fx10{}.

{\em Varying forms of parallelization}:
For each of the core algorithms implemented in \fx10{} and \cx10{}
we present a data parallel implementation.
For five of the core algorithms ({\em BF, DST, BY, DR} and {\em MST})
implemented in \fx10{},
we also provide variations that exploit recursive task parallelism.
Further,
for three of these algorithms ({\em BF, DST} and {\em MST}) we have 
efficient implementations that use clocks (implemented in \cx10{}).


Along with the above discussed variations, we can also vary the runtime 
model (\MP{}, \UP{}, or \SP{} model) by setting the number of places to be a 
divisor of the input (for \MP{} model),  or to the input size 
(for \UP{} model), or to one (for \SP{} model). 

For each of the core distributed algorithms we also present a 
serial implementation in X10. 
The serial implementations do not create any parallel tasks -- they simulate
the behavior of the parallel nodes by serializing their execution in a
predefined order.
Similar to their parallel counterparts, the runtime behavior of the serial
programs can be controlled by varying the underlying runtime model (\MP{}, \UP{}
or \SP{}).
An interesting point to note is that a serial program whose data is distributed
over partitioned global address space mimics a distributed system partly -- where 
accessing remote data is more expensive than accessing the local data.

In summary, we provide a set of 35 (12 in \fx10{}, 11 in \cx10{}, and 12
serial) iterative kernels and 13 (5 in \fx10{}, 3 in \cx10{}, and 5
serial) recursive kernels; 48 kernels in total.


Considering the growing popularity of HJ,
we have also implemented these 48 kernels in 
\fhj{} and \phj{} subsets of HJ (similar to the variations 
provided by \fx10{} and \cx10{}).
Owing to the current limitations of the HJ runtime,
these kernels can only be simulated at a single place.
Thus, we can only realize the \SP{} model of distribution here.




Considering the possibility that in practice these core algorithms may do some
more computation in addition to that specified by the algorithm, all our
kernel benchmarks take an additional option to introduce a user specified
workload in each asynchronous task.
Currently, we present a naive workload function that injects a series of
arithmetic computations (quantity specified by the user at runtime)  to
each asynchronous task in the kernel.
\ifTR{

Our goal is to design a workload that cannot be optimized away by a standard
compiler and further, the workload should not pollute the cache in an overly
adverse manner.
The definition of our workload function is given below.
\begin{tightcode}
long workload(long param) \{
  long j=0;
  for(long i=0; i<loadValue; i++)
    j++;
  return j+param;
\}
\end{tightcode}
Corresponding to each distributed node ({\tt i}), we may create multiple asynchronous tasks;
and each such task invokes the workload function.
A typical invocation of the workload function looks as follows:
\begin{tightcode}
	nval[i] = workload (nval[i] + i);
\end{tightcode}
Note that the value returned by the load function is used again in the next call to the
workload function by the same node.
At the end of the timed computation, 
the sum of the elements of the {\tt nval} array is printed out.
We follow this elaborate procedure to ensure that the workload function is not
optimized away by a compiler.

}
%
We can foresee a workload function with additional characteristics, such
as one that pollutes the L1 / L2 cache, or one that introduces
additional packets in the network and so on. 
The design of such sophisticated workload functions are left as future work.




\subsection{Input generator} 
\label{subsubsec:input}
The input to all the \benchname{} kernels is an abstraction of a distributed system
consisting of the details about its configuration 
(for example, nodes, edges, weights and so on).
\benchname{} comes with a set of input generators that generate inputs specific to 
each kernel benchmark.
Depending on the core algorithm under consideration
each input generator admits a set of options that can be used to tune the 
generated input. 
Some of the common options are the number of nodes in the distributed
graph (referred as the {\em size} of the input), the type of the graph
(complete, sparse and so on), weights of the edges and so on.
Our input generators use a random number generator to generate the details (such as weights, 
adjacency information, unique identifiers of the nodes, and so on) of the distributed graph.
To make the input generation process deterministic, our input generators optionally take
a seed (default value set to the prime number%
\footnote{A set of interesting anecdotes about the number 101 can be found here:
http://primes.utm.edu/curios/page.php?short=101}
$101$).
The users of \benchname{} are required to specify the seed used to generate 
their input;
this can help users to communicate their findings in a more meaningful manner.
Each input generator is serial in nature and is written in Java.

The different types of graphs generated by our input generators depend on
the target algorithm: ring for {\em LCR} and {\em HS}, tree for {\em VC},
and any arbitrary graphs for others.
Considering the typical configurations of trees and arbitrary graphs, our input generators
admit additional options (described below).

{\em Trees}:
We allow three topologies for trees: {\em Star}, {\em Chain} and {\em Random}. 
The last one takes an additional input that specifies the maximum degree for any node.
The choice of Star and Chain as two predefined topologies stems from the behavior of the 
{\em VC} algorithm. 
For a fixed input size, {\em VC} takes the maximum time
for a Star topology and minimum for a Chain.

{\em Arbitrary graphs}: 
Our input generators can generate three types of arbitrary graphs: (i) complete
graphs, (ii) sparse graphs and (iii) random graphs (the edges are
chosen at random).
The limiting cases of the sparse graphs (with edges $n$-$1$ and $n \log n$, where~$n$ 
is the number of nodes) are present as two special options named {\em SP-Min}
and {\em SP-Max}.
To enable the comparison between these two limiting cases, our input generator
ensures that the edge set of {\em SP-Max} variation is a superset of the edge
set of {\em SP-Min}.


\ifTR{
\subsection{Initialization phase} \label{subsubsec:initial}
Initialization stage occurs prior to execution of a kernel.
In this stage, all the data structures pertaining to a node, are fed with the information needed during
execution of the algorithm.
The initialization phase mimics the instance when a node wakes up for the first 
time and joins a network. 
The initialization phases takes a graph as input.
The initialization phase is not timed.
}

\subsection{Output validators} \label{subsec:validate}
Each kernel also consists of an output validator to validate its output. 
The output validator assumes that it has access to the complete input and output
and  may reuse some internal data structures of the main program, for 
efficiency reasons.
\ifTR{
The output verifier for a benchmark program can be enabled by passing the command line option 
"-ver" or "-verify".
}
The output validators are serial in nature and are not timed.

\subsection{Conformance to the key requirements}
We now discuss how the \benchname{} kernels conform to the key
requirements specific to distributed systems (Req\#1~-~\#3).
These kernels are derived from the algorithms that solve some of the core
problems discussed in Figure~\ref{fig:algos-apps}~--~Req\#1.
The \benchname{} kernels cover the important aspects of distributed
systems, such as communication (unicast: {\em LCR}, broadcast:
{\em BY} and {\em DR}, multicast: rest all);
timing (synchronous: {\em DP}, {\em HS}, {\em LCR} and {\em VC}, 
asynchronous: {\em DR} and the recursive kernels of {\em BY}, and partially
synchronous: rest all); and failure: the {\em BY} kernels admit ``nodes"
that may fail (faulty nodes)~--~Req\#2.
Our kernels take as input an abstraction of a set of
(partially) independent nodes and their interconnect. 
By varying the input type we can realize varied interconnects~--~Req\#3.

\section{Evaluation}
\label{s:eval}
We present the characterization of \benchname{} on an IBM cluster
consisting of two hardware nodes%
\footnote{To avoid the confusion between the hardware nodes and the
abstraction of nodes in the input, we explicitly qualify the nodes in the
hardware as ``hardware nodes''. 
We use the generic term ``nodes'' to denote the input nodes.}.
Each hardware node of the cluster has two {Intel~E5-2670 2.6GHz} processors, 
each processor has eight cores and each core can make use of (up to)
two hardware threads.
Thus, we can have up to 64 dedicated hardware threads for our simulations.
Each core has its own local L1 cache that is shared by the two hardware threads.
The two hardware nodes are connected by an FDR10 Infiniband interconnect.
For our simulations we use x10-2.3.0-linux x86 version of X10, jdk1.7.0\_09 version 
of Java and hj-1.3.1 version of HJ.

Our characterization involves, among other things, analyzing the behavior
of the \benchname{} kernels with varying number of available hardware threads (HWTs).
Our hardware configuration  directs the way we increase 
the HWTs for our experiments: 
1/2/4/8 HWTs correspond to one/two/four/eight independent cores on a processor;
16 HWTs correspond to all the cores present in a hardware node;
32 HWTs correspond to all the cores in a hardware node running two hardware threads each;
64 HWTs correspond to 32 HWTs on each of the two hardware nodes.


We use the results of the insightful paper of George et al~\citep{george}
and compute the average running time for our kernels after executing each
of them for 30 times. 
This helps in reducing the noise in the results arising due to many
non-deterministic factors common in a Java based runtime (for example, thread
scheduling, garbage collection and so on).

Considering the fact that many real life network / distributed systems are sparsely connected,
we restrict our evaluation to sparse networks.
%
Specifically, we focus on the limiting cases of sparse inputs:
{\em SP-Max} and {\em SP-Min}.
Similarly for {\em VC}, we use the two corresponding limiting case inputs ({\em Star} and {\em Chain}).
We believe these limiting cases will give us a good understanding of how the
benchmarks may behave for other intermediate inputs.
We refer to these limiting case inputs as \maxin{} and \minin{}, respectively.
However, {\em HS} and {\em LCR} work on ring networks and entertain no such
variations in the network configuration (that is, \maxin{} = \minin{}).


\subsection{Kernel characteristics}

\ifTR{
\begin{figure*}
	\small
        \begin{subfigure}{\textwidth}
	\centering
	\setlength{\tabcolsep}{3pt}
	{
	\begin{tabular}{|c|c|c|c|c|c|c|c|} \hline
	Name 	& \#Static 	& \#Static 	& \#Static 	& \#Dynamic 				& \#Dynamic 					& \multicolumn{2}{c|}{Lines of Code \#}\\\cline{7-8}
	 	& Finish 	& Async 	& mutex  	& Finish 				& Async 					& \fx10{} 			& \fhj{} \\ \hline
	BF 	& 2 		& 2 		& 1 		& $2 \times D$ 				& $2 \times n \times D$ 			& 315 				& 260 \\ \hline
	DST 	& 7 		& 7 		& 3 		& $(3 \times D^2 + 9 \times D)/2$  	& $n \times (3 \times D^2+9\times D)/2$ 	& 490 				& 450 \\ \hline
	BY 	& 3 		& 3 		& 1 		& $(n/8 + 1) (2 \times D + 1)$ 		& $n \times (n/8 + 1) (2 \times D + 1)$ 	& 500 				& 430 \\ \hline
	DR 	& 1 		& 1 		& 0 		& 1 					& n 						& 360 				& 310 \\ \hline
	DS 	& 9 		& 9 		& 3 		& $9 \times (\log^2\Delta\times \log n)$& $9\times n\times(\log^2\Delta\times\log n)$ 	& 630 				& 580 \\ \hline
	KC 	& 10 		& 10 		& 2 		& $5 \times K^2$ 			& $5 \times n \times K^2$ 			& 520 				& 490 \\ \hline
	DP 	& 4 		& 4 		& 1 		& $8 \times D$ 				& $8 \times n \times D$ 			& 460 				& 380 \\ \hline
	HS 	& 5 		& 5 		& 0 		& $\log n \times (6 \times n+1) + 1$ 	& $n \times (\log n \times (6 \times n+1) + 1)$	& 460 				& 430 \\ \hline
	LCR 	& 3 		& 3 		& 0 		& $2 \times n+1$ 			& $2 \times n^2 + n$ 				& 250 				& 210 \\ \hline
	MIS 	& 5 		& 5 		& 3 		& $(3 \log_{4/3}m + 1)\times 4 + 1$ 	& $4 \times n \times (3 \log_{4/3}m + 1) + n$ 	& 380 				& 320\\ \hline
	MST 	& 15 		& 15 		& 4 		& $3 \times D^2 + 11 \times D$ 		& $3 \times n \times D^2 + 11 \times n \times D$& 900 				& 810 \\ \hline
	VC 	& 5 		& 5 		& 0 		& $2 \log^{*}n + 9$ 			& $2 \times n \log^{*}n + 9 \times n$ 		& 430 				& 375 \\ \hline

	\end{tabular}
	}
	\caption{{\bf \fx10{} and \fhj{} Kernels characteristics.}}
	\label{fig:fx10-characteristics}
	
	\end{subfigure}%
		
	\begin{subfigure}{\textwidth}
			\resizebox{\textwidth}{!} 
			{
			\begin{tabular}{|c|c|c|c|c|c|c|c|c|c|} \hline
			Name & \#Static & \#Static & \#Static & \#Static & \#Dynamic & \#Dynamic & \#Dynamic & \multicolumn{2}{|c|}{Lines of Code \#}\\\cline{9-10}
				& \finish{} & \async{} & barrier  & mutex  & \finish{} & \async{} & barrier  & \cx10{} & \phj{} \\ \hline
			BF & 1 & 1 & 1 & 1 & $D$ & $n \times D$ & $n \times D$ & 300 & 250 \\ \hline
			DST & 5 & 5 & 2 & 3 & $D^2 + 3 \times D$  & $n \times (D^2 + 3 \times D)$ & $(D^2 + 3 \times D)/2$ & 480 & 430 \\ \hline
			BY & 2 & 2 & 1 & 1 & $(n/8 + 1) \times (D + 1)$ & $n \times (n/8 + 1)  \times  (D + 1)$ & $D \times (n/8+1) $ & 500 & 420  \\ \hline
			DS & 1 & 1 & 8 & 3 & $\log^2 \Delta  \times \log n$ & $n \times (\log^2 \Delta  \times \log n)$ & $8 \times (\log^2 \Delta  \times \log n)$ & 530 & 450 \\ \hline
			KC & 7 & 7 & 3 & 2 & $2 \times K^2 + 3 \times K$ & $2 \times n \times K^2 + 3 \times n \times K$ & $3 \times K^2 - 3 \times K$ & 500 & 470 \\ \hline
			DP & 1 & 1 & 3 & 1 & $2 \times D$ & $2 \times n \times D$ & $6 \times D$ & 435 & 340 \\ \hline
			HS & 3 & 3 & 2 & 0 & $\log n  \times (2 \times n+1) + 1$ & $n \times (\log n *(2 \times n+1) + 1)$ & $4 \times n \log n $ & 440 & 425 \\ \hline
			LCR & 2 & 2 & 1 & 0 & $n+1$ & $n^2 + n$ & $n+1$ & 240 & 200 \\ \hline
			MIS & 2 & 2 & 3 & 3 & $(3 \log_{4/3}m + 1) + 1$ & $n \times (3 \log_{4/3}m + 1) + n$ & $3 \times (3 \log_{4/3}m + 1)$ & 360 & 290 \\ \hline
			MST & 8 & 8 & 9 & 4 & $2 \times D^2 + 6 \times D$ & $2 \times n \times D^2 + 6 \times n \times D$ & $2 \times n \times D^2 + 7 \times n \times D$ & 840 & 750 \\ \hline
			VC & 3 & 3 & 2 & 0 & $\log^{*}n + 6$ & $n \log^{*}n + 6 \times n$ & $\log^{*}n + 3$ & 395 & 350 \\ \hline

			\end{tabular}
			}
			\caption{{\bf \cx10{} and \phj{} Kernels characteristics.}}
			\label{fig:cx10-characteristics}
			
			\end{subfigure}
       
\caption{{\bf Static and Dynamic characteristics of the \benchname{} kernels }
Here $D$ represents the diameter of the graph, $n$ represents the number of
nodes, $\Delta$ represents the maximum degree of a graph, $K$ represents the
maximum size of a committee. For each program number of static
\async{}s = number of static \finish{}, and number of dynamic \async{}s = $n
\times$ number of dynamic \finish{}.}
\label{fig:characteristics}
\end{figure*}
}

In this section, we discuss some of the static and dynamic characteristics of 
our kernels.
Figure~\ref{fig:fx10-characteristics} and Figure~\ref{fig:fx10-rec-characteristics} 
present these characteristics for the
iterative and recursive kernels, respectively, written in \fx10{} and \fhj{}.
Similarly, Figure~\ref{fig:cx10-rec-characteristics} and
Figure~\ref{fig:cx10-characteristics} present these characteristics for the
recursive and iterative kernels, respectively, written in \cx10{} and \phj{}.
In these tables, \amutex{} is used as a generic name referring to mutex operations -- {\tt atomic} construct in X10 and {\tt isolated} construct in HJ.
Similarly,
\abarrier{} is used as a generic name referring to barrier operations -- {\tt Clock.advanceAll()} construct in X10 and {\tt next} in HJ.
We use the following abbreviation: a) "\#Static/Dynamic \afinish{}" to represent number of
static/dynamic \finish{} operations, 
b) {\tt \#Comm} to represent the number of remote communications (excluding the barrier operations).

It can be seen that all the kernels in \benchname{} are relatively small in size; 
their sizes vary approximately between 200 to 900 lines.
Further the number of static \finish{}, mutex and barrier statements are quite small.
We discuss the characteristics of the iterative and recursive kernels
separately.

\subsubsection{Iterative kernels}
The number of dynamic \finish{} and barrier statements vary as a function of the actual input.
The number of static \async{} statements matches the number of static \finish{} statements, 
while the number of dynamic \async{}s is $n$ times the number of dynamic \finish{} statements.
Unlike the counts of the dynamic \finish{} statements, the counts of
the dynamic mutex and remote communications may depend on the structure of the
input graph.
Hence for these operations, we present the runtime characteristics (of programs written in \fx10{} and \cx10{}) by comparing them for two
specific inputs (\maxin{} with 64 nodes and \minin{} with 64 nodes, both
run on 64 runtime places),
in Figure~\ref{fig:fx10-comm-characteristics}.
\begin{figure}[H]
\small
\begin{subfigure}{\columnwidth}
\centering
\begin{tabular}{|c|c|c|c|c|c|} \hline
Name 		& \multicolumn{2}{c|}{\#Lines of Code}	& \multicolumn{2}{c|}{\#Static}	& \#Dynamic \afinish{} 			\\ \cline{2-5}
	 	& \fx10{} 			& \fhj{}&  \afinish{}	& \amutex{}  	&  					\\ \hline
{\em BF} 	& 330 				& 320 	& 2 		& 1 		& $2 \times D$ 				\\ \hline
{\em DST} 	& 510 				& 500 	& 7 		& 3 		& $(3 \times D^2 + 9 \times D)/2$  	\\ \hline
{\em BY} 	& 510 				& 460 	& 3 		& 1 		& $(n/8 + 1) (2 \times D + 1)$ 		\\ \hline
{\em DR} 	& 370 				& 355 	& 1 		& 0 		& 1 					\\ \hline
{\em DS} 	& 650 				& 600 	& 9 		& 2 		& $9 \times (\log^2\Delta\times \log n)$\\ \hline
{\em KC} 	& 490 				& 490 	& 10 		& 2 		& $5 \times K^2$ 			\\ \hline
{\em DP} 	& 440 				& 395 	& 4 		& 1 		& $8 \times D$ 				\\ \hline
{\em HS} 	& 440 				& 435 	& 5 		& 0 		& $\log n \times (6 \times n+1) + 1$ 	\\ \hline
{\em LCR} 	& 260 				& 245 	& 3 		& 0 		& $2 \times n+1$ 			\\ \hline
{\em MIS} 	& 380 				& 330	& 5 		& 3 		& $(3 \log_{4/3}m + 1)\times 4 + 1$ 	\\ \hline
{\em MST} 	& 880 				& 790 	& 15 		& 4 		& $\log n \times (3 \times D + 11)$ 	\\ \hline
{\em VC} 	& 420 				& 395 	& 5 		& 0 		& $2 \log^{*}n + 9$ 			\\ \hline
\end{tabular}
\caption{{\fx10{} and \fhj{} iterative kernels.}}
\label{fig:fx10-characteristics}
\end{subfigure}	

\begin{subfigure}{\columnwidth}\small
\centering
\begin{tabular}{|c|c|c|c|c|c|c|c|c|c|} \hline
Name	& \multicolumn{2}{c|}{\#Lines of Code}      	& \multicolumn{2}{c|}{\#Static}		& \multicolumn{2}{c|}{\#Lines of Code}      	& \multicolumn{3}{c|}{\#Static}				\\ \cline{2-10}	
		&\hspace*{-0.03in}\fx10{}\hspace*{-0.03in} 			&\hspace*{-0.03in} \fhj{} \hspace*{-0.03in}     	&\hspace*{-0.03in} \afinish{}\hspace*{-0.03in} 		&\hspace*{-0.03in} \amutex{}\hspace*{-0.03in}  	&\hspace*{-0.03in}\cx10{}	\hspace*{-0.03in}& \hspace*{-0.03in}\phj{}  \hspace*{-0.03in}  			& \hspace*{-0.03in}\afinish{} \hspace*{-0.03in}		& \hspace*{-0.03in}\abarrier{} \hspace*{-0.03in} 	& \hspace*{-0.03in}\amutex{}  \hspace*{-0.03in}	\\ \hline
{\em BF} 	& 275 				& 275	      	& 2 			& 1 		& 270 		& 270	    			& 1 			& 1 		& 1 		\\ \hline
{\em DST} 	& 480 				& 475	      	& 6 			& 3 		& 475 		& 470	    			& 5 			& 1 		& 3 		\\ \hline
{\em MST} 	& 705 				& 630	      	& 11 			& 3 		& 675 		& 590	    			& 4 			& 7 		& 3		\\ \hline
{\em BY} 	& 425 				& 440	      	& 3 			& 2 		& \multicolumn{5}{c|}{--}										\\ \hline
{\em DR} 	& 375 				& 360	      	& 3 			& 0 		& \multicolumn{5}{c|}{--}										\\ \hline
\end{tabular}
\caption{{Recursive kernels. }}
\label{fig:fx10-rec-characteristics}
\label{fig:cx10-rec-characteristics}
\end{subfigure}

\begin{subfigure}{\columnwidth} \small
\centering
\begin{tabular}{|c|c|c|c|c|c|c|c|} \hline
Name 		& \multicolumn{2}{c|}{\#Lines of Code}      	& \multicolumn{3}{c|}{\#Static}			& \multicolumn{2}{c|}{\#Dynamic} 							\\\cline{2-8}
		& \cx10{} 			& \phj{}      	& \afinish{} 	& \abarrier{}  	& \amutex{}  	& \afinish{} 				& \abarrier{}  					\\ \hline
{\em BF} 	& 330 				& 310	      	& 1 		& 1 		& 1 		& $D$ 					& $D$	 					\\ \hline
{\em DST}	& 500 				& 490	      	& 5 		& 2 		& 3 		& $D^2+3 \times D$  			& $\frac{D^2+3 \times D}{2}$			\\ \hline
{\em BY} 	& 505 				& 455	      	& 2 		& 1 		& 1 		& $(D+1)\times$  			& $D \times$ 					\\ 
		&				&	   	&		&		&		& $(\frac{n}{8}+1)$			& $(\frac{n}{8}+1)$				\\ \hline
{\em DS} 	& 580 				& 510	      	& 1 		& 8 		& 2 		& $\log^2 \Delta$  			& $8 \times \log^2 \Delta$  			\\
		&				&		&		&		&		& $\times \log n$			& $\times \log n$				\\ \hline
{\em KC} 	& 480 				& 470	      	& 7 		& 3 		& 2 		& $2 \times K^2 +$  			& $3 \times K^2 +$  				\\ 
		&				&		&		&		&		& $3 \times K$				& $3 \times K$					\\ \hline
{\em DP} 	& 430 				& 375	      	& 1 		& 3 		& 1 		& $2 \times D$ 				& $6 \times D$ 					\\ \hline
{\em HS} 	& 430 				& 420	      	& 3 		& 2 		& 0 		& $1 + \log n$   			& $4 \times n \log n $ 				\\
		&				&		&		&		&		& $\times (2 \times n+1)$		&						\\ \hline
{\em LCR} 	& 250 				& 240	      	& 2 		& 1 		& 0 		& $n+1$ 				& $n+1$						\\ \hline
{\em MIS} 	& 365 				& 310	      	& 2 		& 3 		& 3 		& $1+(1+$  				& $3 \times (1+$ 				\\
		&				&		&		&		&		& $3 \log_{4/3}m)$			& $3 \log_{4/3}m)$				\\ \hline
{\em MST} 	& 850 				& 760	      	& 8 		& 9 		& 4 		& $(2 \times D + 6)$ 			& $(2 \times D + 7)$				\\
		&				&		&		&		&		& $\times \log n$			& $\times \log n$				\\ \hline
{\em VC} 	& 410 				& 390	      & 3 		& 2 		& 0 		& $\log^{*}n + 6$ 			& $\log^{*}n + 3$ 				\\ \hline
\end{tabular}
\caption{{Iterative \cx10{} and \phj{} kernels.}}
\label{fig:cx10-characteristics}
			
\end{subfigure}
\caption{Static characteristics of \benchname{} kernels; $D$ represents the diameter, $n$ represents 
the number of nodes, $\Delta$ represents the maximum degree, and $K$ represents 
the required maximum committee size.}
\end{figure}
We avoid presenting the numbers for \fhj{} and \phj{} separately as the number of mutex operations
match exactly that of \fx10{} and \cx10{}, respectively, and 
since \fhj{} and \phj{} kernels run in the context of \SP{} model
they do not involve any remote communication.

Note that for a given input, the number of static (and dynamic) mutex operations is same for 
both \fx10{} and \cx10{} kernels.
This is because these two mainly differ in the synchronization primitives they use 
(see Section~\ref{s:implement}).
For the same reason, the \cx10{}  kernels have fewer number of static and
dynamic \finish{} (and \async{}) operations compared to the \fx10{}
kernels.

\begin{figure*}
\small
\centering
\begin{tabular}{|c|c|c|c|c|c|c|} \hline

Name 		& \multicolumn{2}{c|}{{\tt \#Comm (FA)}}	& \multicolumn{2}{c|}{\#\amutex{}} 	& \multicolumn{2}{c|}{{\tt \#Comm (FAC)}}	\\\cline{2-7}
     		& \minin{} 	& \maxin{} 		& \minin{} 	& \maxin{}  		& \minin{} 	& \maxin{}  				\\ \hline
{\em BF} 	& 1009 		& 1145 			& 126 		& 766 			& 568 		& 956	        	      		\\ \hline
{\em DST} 	& 7500 		& 3446 			& 343 		& 1348	 		& 5232 		& 2816	        	      		\\ \hline
{\em BY} 	& 46386 	& 97299 		& 43K 		& 95K 			& 44874		& 96291	        	      		\\ \hline
{\em DR} 	& 20587  	& 16204 		& 0 		& 0 			& --		& --	        	      		\\ \hline
{\em DS} 	& 8020 		& 31118 		& 124 		& 327			& 6193 		& 27401         	      		\\ \hline
{\em KC} 	& 5082 		& 12504 		& 2243	 	& 9666 			& 4125 		& 11556	        	      		\\ \hline
{\em DP} 	& 9996 		& 12703 		& 2625 		& 8472 			& 5649 		& 10624	        	      		\\ \hline
{\em HS} 	& 12223 	& 12223 		& 0 		& 0 			& 8255 		& 8255	        	      		\\ \hline
{\em LCR} 	& 25373 	& 25373 		& 0 		& 0 			& 9229 		& 9229	        	      		\\ \hline
{\em MIS} 	& 1463 		& 3204 			& 1894		& 3518			& 1274 		& 2637 	        	      		\\ \hline
{\em MST} 	& 8890 		& 19154 		& 497 		& 469 			& 6055 		& 13547	        	      		\\ \hline
{\em VC} 	& 1008 		& 1208 			& 0 		& 0 			& 756 		& 890	        	      		\\ \hline
\end{tabular}
\caption{{\fx10{} \& \cx10{} dynamic communication and mutex operations}; input size = 64 nodes.}
\label{fig:fx10-comm-characteristics}
\end{figure*}

{\bf Analysis of dynamic mutex operations and communication}:
As shown in Figure~\ref{fig:fx10-comm-characteristics},
the number of mutex operations for \maxin{} is consistently higher than
that of \minin{}, except in case of {\em MST}.
The {\em MST} kernel has an interesting property that
the number of mutex operations is guaranteed not to grow as we introduce
some additional edges and corresponding unique weights.
Thus the number of dynamic mutex operations for the \maxin{} 
is not higher than that for \minin{}.
The kernels {\em DR}, {\em HS, LCR}, and {\em VC} have no mutex operations and it is reflected in 
Figures~\ref{fig:fx10-characteristics}, \ref{fig:cx10-characteristics}, and~\ref{fig:fx10-comm-characteristics}.


An interesting point about {\em MIS} is that the amount of dynamic
communication is less than the number of mutex operations.
This is because in {\em MIS}, majority of remote communication operations involve mutex
operation, but the other way round is not true.

Note that, except for {\em DST} and {\em DR} kernels, rest all the kernels have 
higher amount of communication for \maxin{} compared to \minin{}.
One characteristic difference between \maxin{} and \minin{} is that the latter
increases the diameter of the graph and hence impacts the algorithms where an 
increase in diameter causes an increase in rounds.
For {\em DST} and {\em DR} as the number of rounds increases, the number of
messages (amount of remote communication)  also increases.

\begin{figure}[H]
	\small
        \begin{subfigure}{0.33\textwidth}
                \includegraphics[height=0.645\textwidth,width=\textwidth]{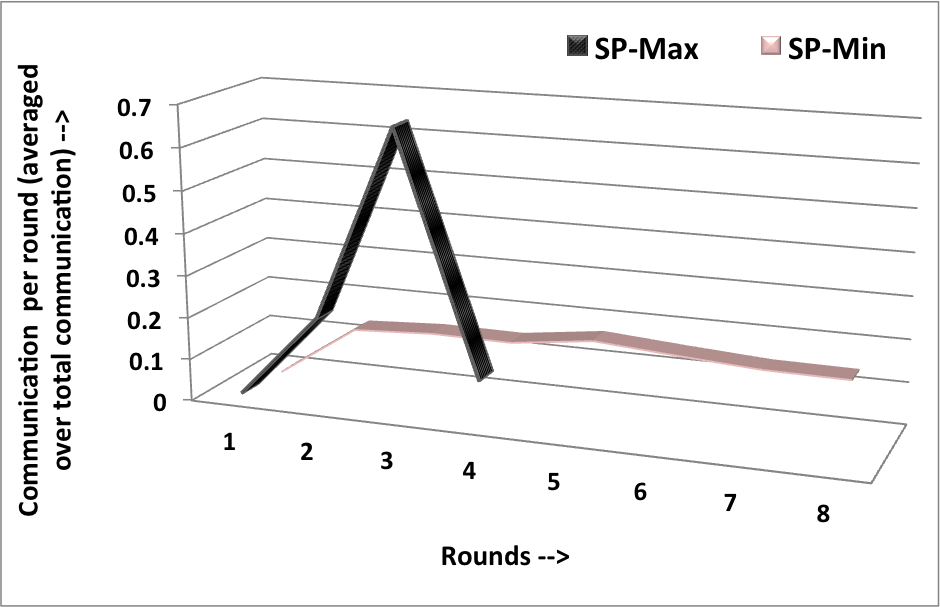}
                \caption{{\em BF}} 
                \label{fig:bf_pr}
        \end{subfigure}%
        \begin{subfigure}{0.33\textwidth}
                \includegraphics[height=0.645\textwidth,width=\textwidth]{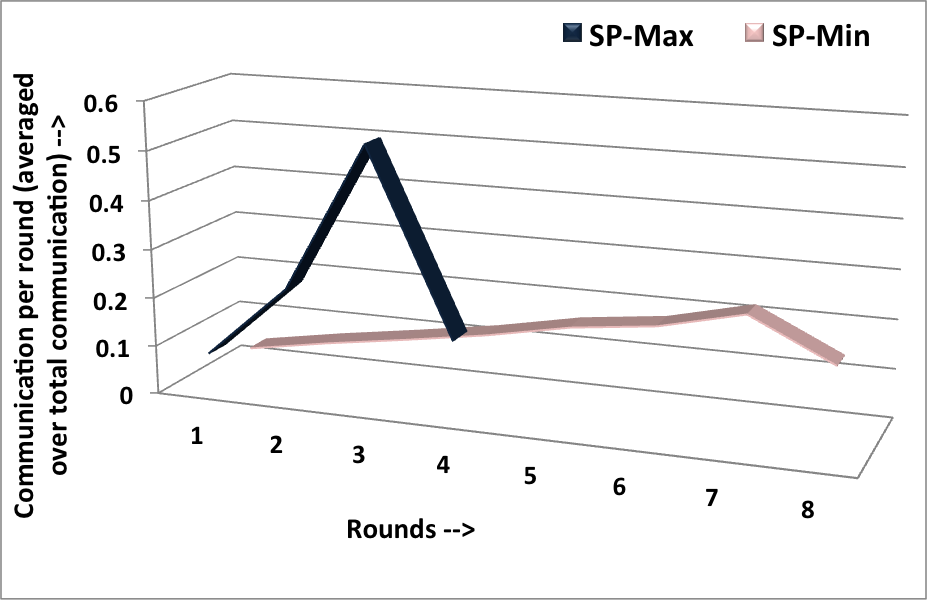}
                \caption{{\em DST}} 
                \label{fig:dst_pr}
        \end{subfigure}%
        \begin{subfigure}{0.33\textwidth}
                \includegraphics[height=0.64\textwidth,width=\textwidth]{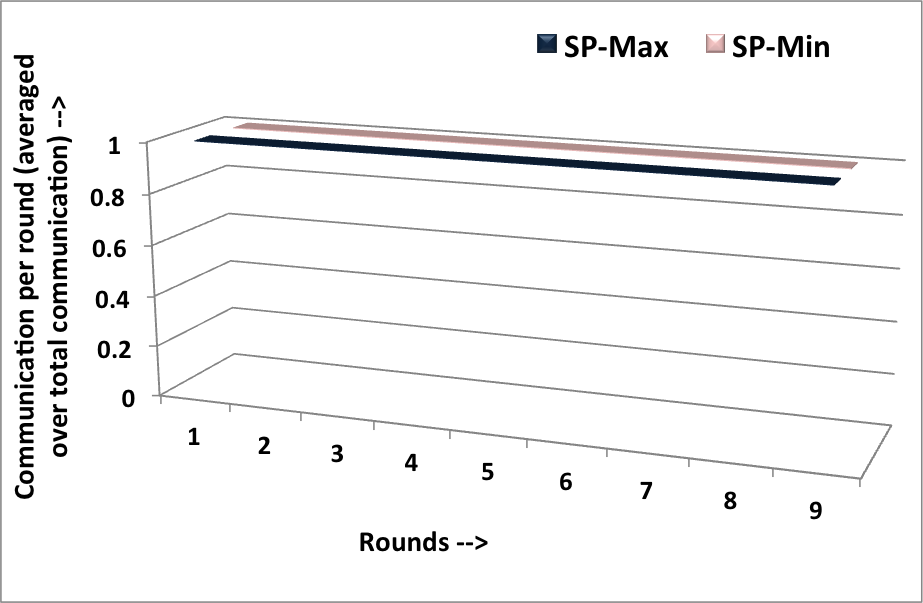}
                \caption{{\em BY}} 
                \label{fig:by_pr}
        \end{subfigure}%

        \begin{subfigure}{0.333\textwidth}
                \includegraphics[height=0.645\textwidth,width=\textwidth]{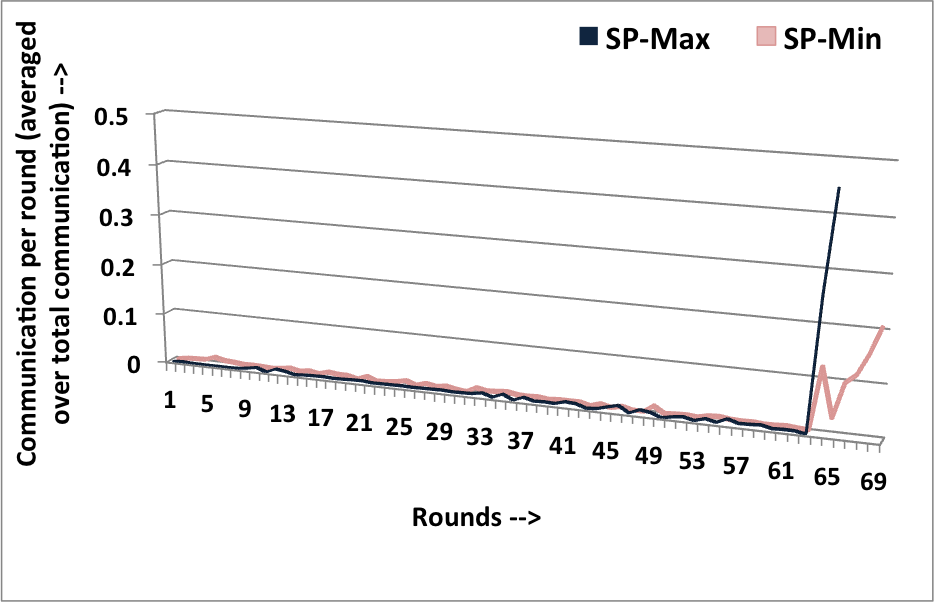}
                \caption{{\em DR}} 
                \label{fig:dr_pr}
        \end{subfigure}%
        \begin{subfigure}{0.333\textwidth}
                \includegraphics[height=0.645\textwidth,width=\textwidth]{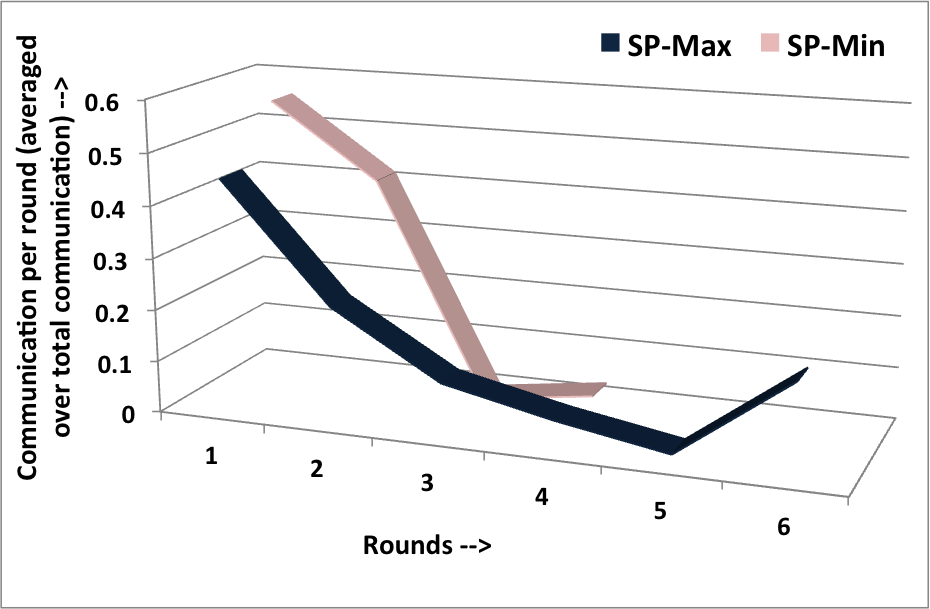}
                \caption{{\em DS}} 
                \label{fig:ds_pr}
        \end{subfigure}%
        \begin{subfigure}{0.333\textwidth}
                \includegraphics[height=0.645\textwidth,width=\textwidth]{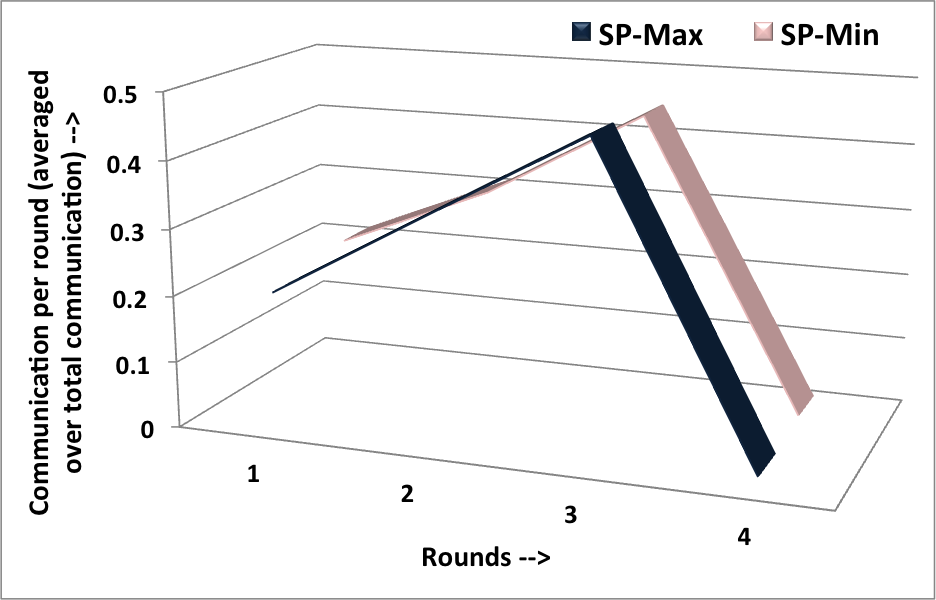}
                \caption{{\em KC}} 
                \label{fig:kc_pr}
        \end{subfigure}%

        \begin{subfigure}{0.333\textwidth}
                \includegraphics[height=0.645\textwidth,width=\textwidth]{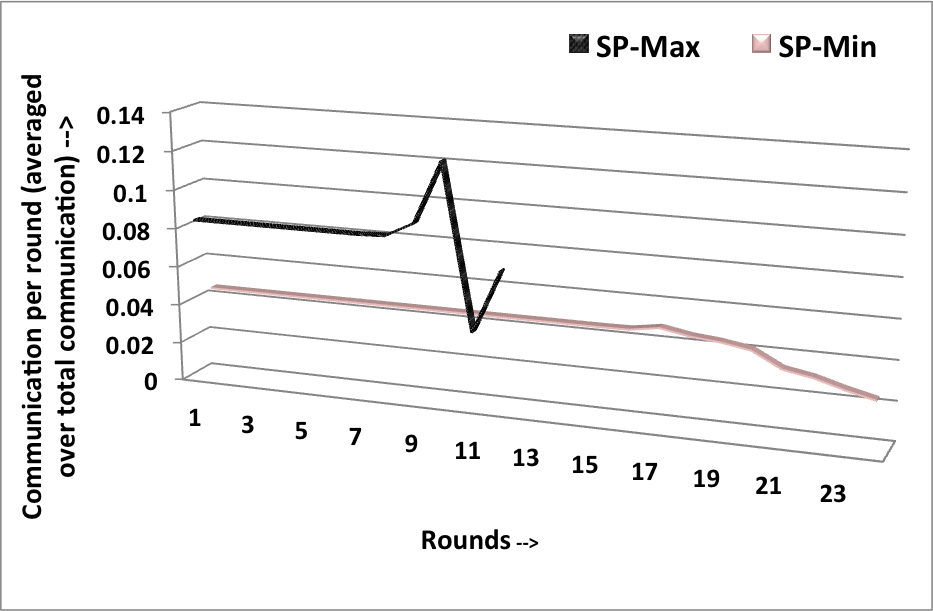}
                \caption{{\em DP}} 
                \label{fig:dp_pr}
        \end{subfigure}%
        \begin{subfigure}{0.333\textwidth}
                \includegraphics[height=0.645\textwidth,width=\textwidth]{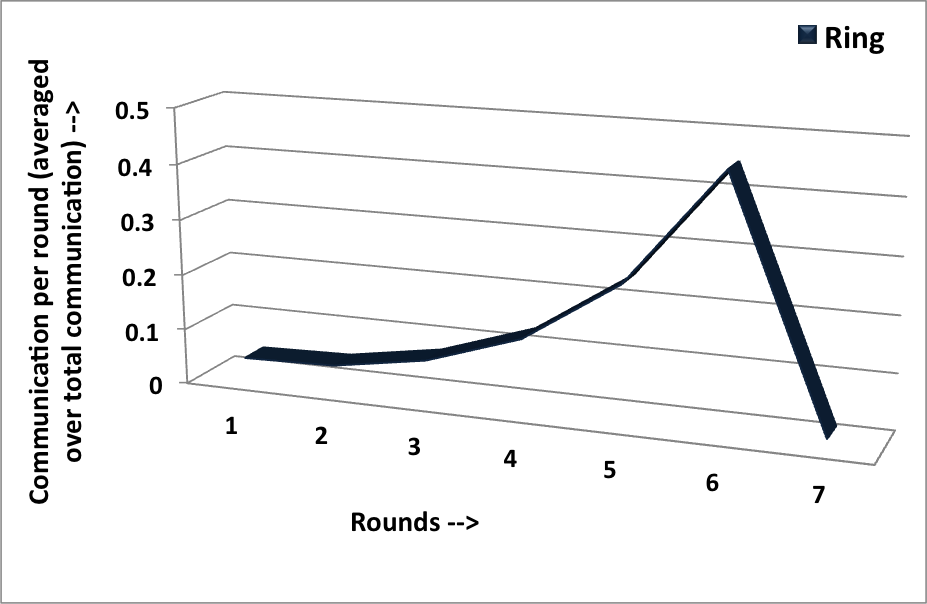}
                \caption{{\em HS}} 
                \label{fig:hs_pr}
        \end{subfigure}%
        \begin{subfigure}{0.333\textwidth}
                \includegraphics[height=0.645\textwidth,width=\textwidth]{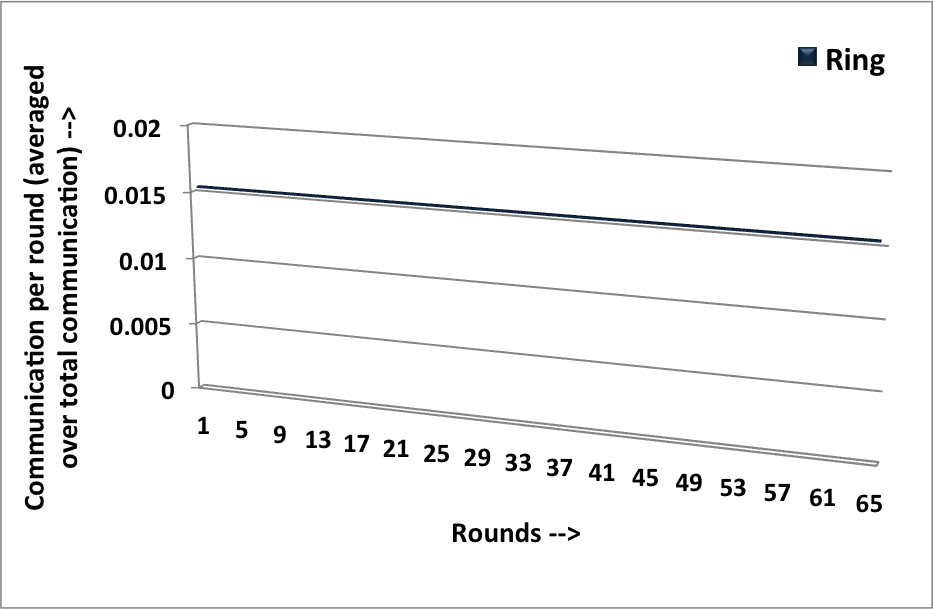}
                \caption{{\em LCR}} 
                \label{fig:lcr_pr}
        \end{subfigure}%

	\begin{subfigure}{0.333\textwidth}
                \includegraphics[height=0.645\textwidth,width=\textwidth]{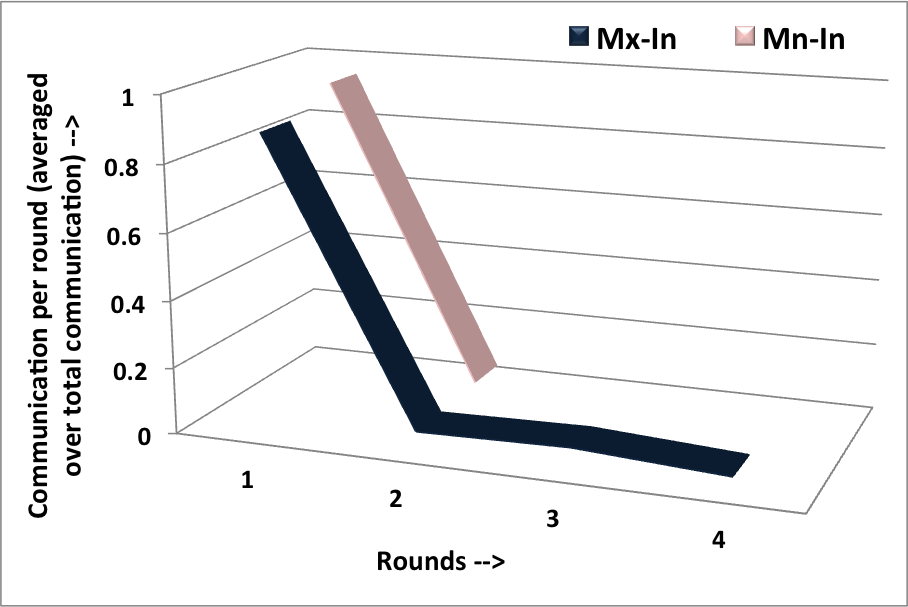}
                \caption{{\em MIS}} 
                \label{fig:mis_pr}
        \end{subfigure}%
	\begin{subfigure}{0.333\textwidth}
                \includegraphics[height=0.645\textwidth,width=\textwidth]{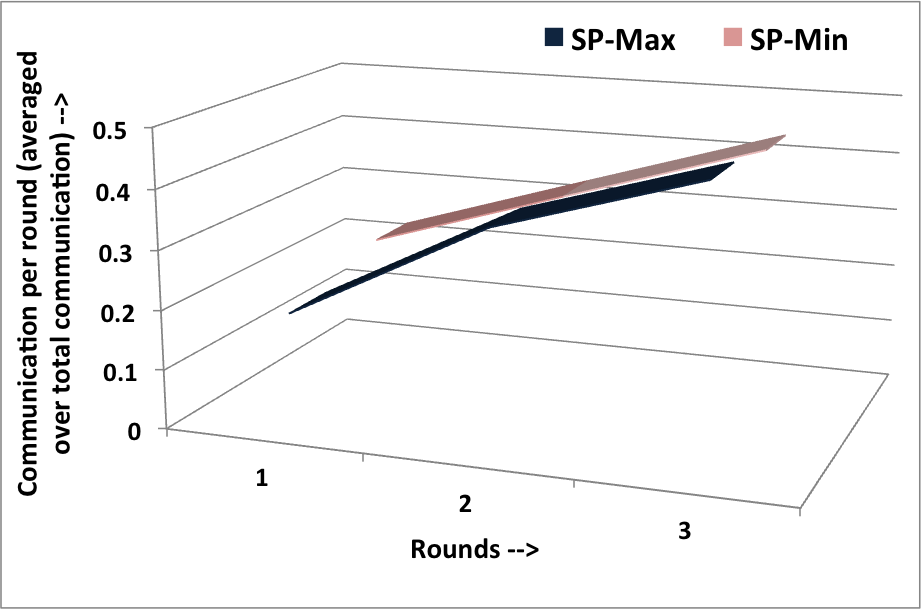}
                \caption{{\em MST}} 
                \label{fig:mst_pr}
        \end{subfigure}%
        \begin{subfigure}{0.333\textwidth}
                \includegraphics[height=0.645\textwidth,width=\textwidth]{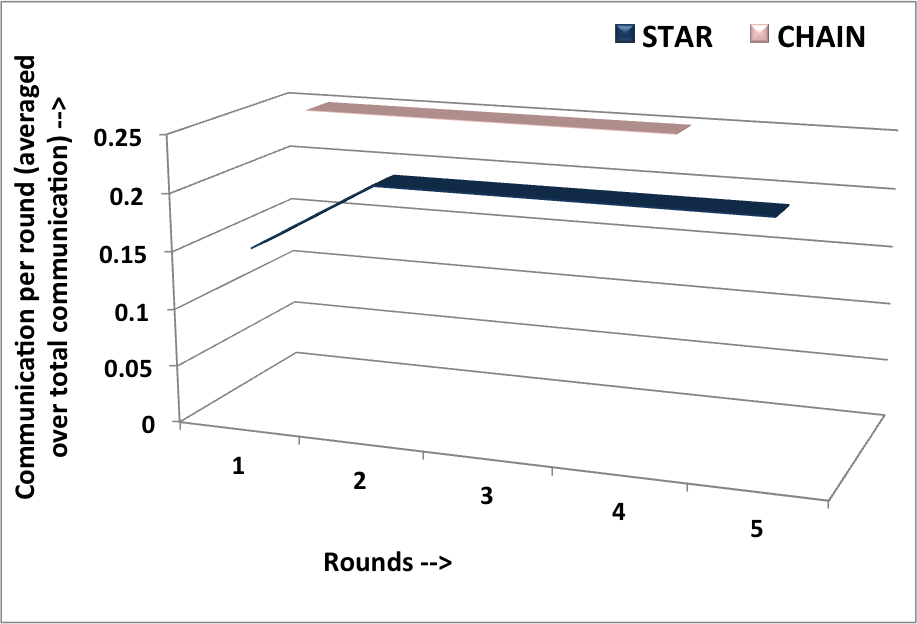}
                \caption{{\em VC}} 
                \label{fig:vc_pr}
        \end{subfigure}%

        \caption{ \cx10{} plots for dynamic communication per round; input size = 64 nodes, \#~clusters = 64.}
        \label{fig:cx10-per-round-comm-characteristics}
\end{figure}

{\bf Communication distribution}
Figure~\ref{fig:cx10-per-round-comm-characteristics} shows the amount of remote
communication occurring in each round, for the \cx10{} kernels%
\footnote{Considering the case that we do not have a separate \cx10{} version
for {\em DR}, we use the plot of the corresponding \fx10{} version here.};
for the sake of illustration we set the input size to 64 nodes and present the
results for two types of inputs: \maxin{} and \minin{}.
For {\em LCR} and {\em HS} we show only one curve as in the context of ring network
\maxin{} = \minin{}.

The behavior of {\em BY}, {\em KC} and {\em MST} for \maxin{} and
\minin{} are quite similar.
Note that in {\em BY} for a specific input the amount of 
communication in each round is equal, but the communication 
per round in \minin{} less than \maxin{}.
In case of {\em BF}, {\em DST}, {\em DR}, {\em DS} and {\em DP}, compared to \minin{}, the algorithm terminates in fewer
rounds in case of \maxin{}.
However {\em MIS} and {\em VC} exhibit a contrasting behavior.
In {\em MIS}, compared to \minin{} where each node has fewer
neighbors than \maxin{}, in each round fewer nodes are
added to the maximal-independent-set in case of \maxin{};
thus increasing the number of rounds.
In {\em VC}, as compared to \minin{} where the algorithm requires
exactly one round to make the graph six colored, in \maxin{} the
number of rounds ($\geq 1$) depends on the input.
The shift-down operation (Section~\ref{s:back}) on the other hand always
takes three rounds to finish (irrespective of the input).
%
For lack of space, we omit the communication distribution plots for the \fx10{} and
recursive kernels.

\begin{figure}[t]
\small
\begin{subfigure}{\columnwidth}
\centering
\begin{tabular}{|c|c|c|c|c|c|c|c|c|} \hline
Name 		& \multicolumn{2}{c|}{{\tt \#Async}}	& \multicolumn{2}{c|}{\#\afinish{}} 	& \multicolumn{2}{c|}{{\tt \#Comm}}	& \multicolumn{2}{c|}{\#\amutex{}}	\\\cline{2-9}
     		& \minin{} 	& \maxin{} 		& \minin{} 	& \maxin{}  		& \minin{} 	& \maxin{}		& \minin{}	& \maxin{}		\\ \hline
{\em BF} 	& 252		& 2824			& 128		& 240			& 317		& 2945			& 317		& 2945			\\ \hline 
{\em DST} 	& 2201 		& 1092 			& 256		& 151			& 2721		& 2569			& 189		& 1279			\\ \hline
{\em BY}	& 73K		& 435K			& 36K		& 36K			& 74K  		& 442K   		& 109K  	& 478K   		\\ \hline	
{\em DR}	& 3476		& 3699			& 375		& 279			& 12469		& 12373			& 0		& 0			\\ \hline		
{\em MST}	& 1872		& 1870			& 258		& 345			& 15341		& 16460			& 255		& 255			\\ \hline
\end{tabular}

\caption{{\fx10{} recursive kernel characteristics; input size = 64 nodes.}}
\label{fig:rec-fx10-comm-characteristics}
\end{subfigure}

\begin{subfigure}{\columnwidth}\small
\centering
\begin{tabular}{|c|c|c|c|c|} \hline
Name 		& \multicolumn{2}{c|}{{\tt \#Async}}	& \multicolumn{2}{c|}{\#\afinish{}} 	\\\cline{2-5}
     		& \minin{} 	& \maxin{} 		& \minin{} 	& \maxin{}  		\\ \hline
{\em BF} 	& 126		& 1615			& 64		& 140			\\ \hline 
{\em DST} 	& 1689		& 836			& 248		& 147			\\ \hline
{\em MST}	& 528		& 526			& 237		& 324			\\ \hline
\end{tabular}

\begin{tabular}{|c|c|c|c|c|c|c|} \hline
Name 		& \multicolumn{2}{c|}{{\tt \#Comm}}	& \multicolumn{2}{c|}{\#\abarrier{}}  & \multicolumn{2}{c|}{\#\amutex{}} \\\cline{2-7}
     		& \minin{} 	& \maxin{}		& \minin{}	& \maxin{}	& \minin{}	& \maxin{}		\\ \hline
{\em BF} 	& 317		& 3371			& 64		& 140		& 317		& 3371			\\ \hline 
{\em DST} 	& 2217		& 2317			& 8		& 4		& 189		& 1279			\\ \hline
{\em MST}	& 14026		& 15159			& 21		& 21		& 255		& 255			\\ \hline
\end{tabular}

\caption{{\cx10{} recursive kernel characteristics; input size = 64 nodes.}}
\label{fig:rec-cx10-comm-characteristics}
\end{subfigure}
\caption{Recursive kernels - runtime characteristics.}
\label{fig:rec-all}
\end{figure}

\subsubsection{Recursive kernels}
For our recursive kernels,
Figure~\ref{fig:fx10-rec-characteristics} presents the static characteristics, and
Figures~\ref{fig:rec-fx10-comm-characteristics} and~\ref{fig:rec-cx10-comm-characteristics}
present the runtime characteristics,
for inputs \minin{} and \maxin{}.
For the most part,
the comparative behavior (between \minin{} and \maxin{}) displayed by these recursive
kernels is similar to their iterative counterparts.
A few points of interest:
(i) the recursive {\em BY} kernel has more mutex operations than communication.
This is because in the recursive kernel, majority of remote communication operations involve mutex
operation, but the other way round is not true.
(ii) in case of recursive {\em DST} kernel,
the reduction in the amount of communication between the \fx10{} and \cx10{} versions
is directly impacted by the number of remote task creation operations present in the program:
\cx10{} has comparatively fewer remote task creation operations than \fx10{}.
(iii) in case of {\em MST} the number of mutex operations (in \fx10{} and \cx10{}) and 
barriers (in \cx10{}) are equal for both \maxin{} and \minin{} input, as they are 
independent of the structure and type of input.
However, the number of {\em async} and {\em finish} operations depends on the 
exact structure of the graph and the edge weights;
thus making it hard to correlate the numbers for these two operations with the input types.
%


\subsection{Performance analysis}
%

In this section we study the 
effect of three {\em key} parameters on the behavior of \benchname{} kernels.
These key parameters are:
(a) number of available hardware threads (HWTs), (b) input size (denoting the number of
nodes), and (c) number of node clusters%
\footnote{We assume that the nodes are distributed equally among all the
clusters.}.
Variations in the number of node clusters are achieved by
varying the number of runtime places in the \fx10{} and \cx10{}  kernels.
We study the effect of these parameters both in isolation (by varying only one 
parameter and fixing the rest) and in
conjunction with each other (by varying two or three parameters at a time
and fixing the rest). 
Varying multiple parameters at the same time may lead to an overly large
number of experimental points. 
We handle this situation by varying these parameters in ``sync'' (all the varying
parameters get the same value, for a given evaluation point).
For the purpose of this study, we set the input type to \maxin{}.
Later (Section~\ref{sss:vit}),  we also analyze the 
effect of input type (\maxin{} and \minin{}) on
the behavior of the \benchname{} kernels.

{\em Limits for the key parameters of our study}: 
We vary the input size between 8 to 512 nodes. The execution times are
too insignificant for inputs smaller than 8 nodes and it takes too long
(of the order of several days) to execute programs with inputs larger
than 512 nodes.
We vary the number of HWTs between 1 to 64 (limited by the experimental
system at hand) when using the X10 runtime. 
The absence of a distributed HJ runtime limits the maximum number of HWTs,
for running our HJ based kernels, to 32 on our hardware.
We vary the number of clusters between 1 to 64;
our version of X10 runtime throws a runtime error (ERRNO 104 - connection
reset by peer) if we request more than 64 places on our machine.



\ifTR{
\begin{figure*}
        \begin{subfigure}{0.4\textwidth}
                \includegraphics[height=0.8\textwidth,width=\textwidth]{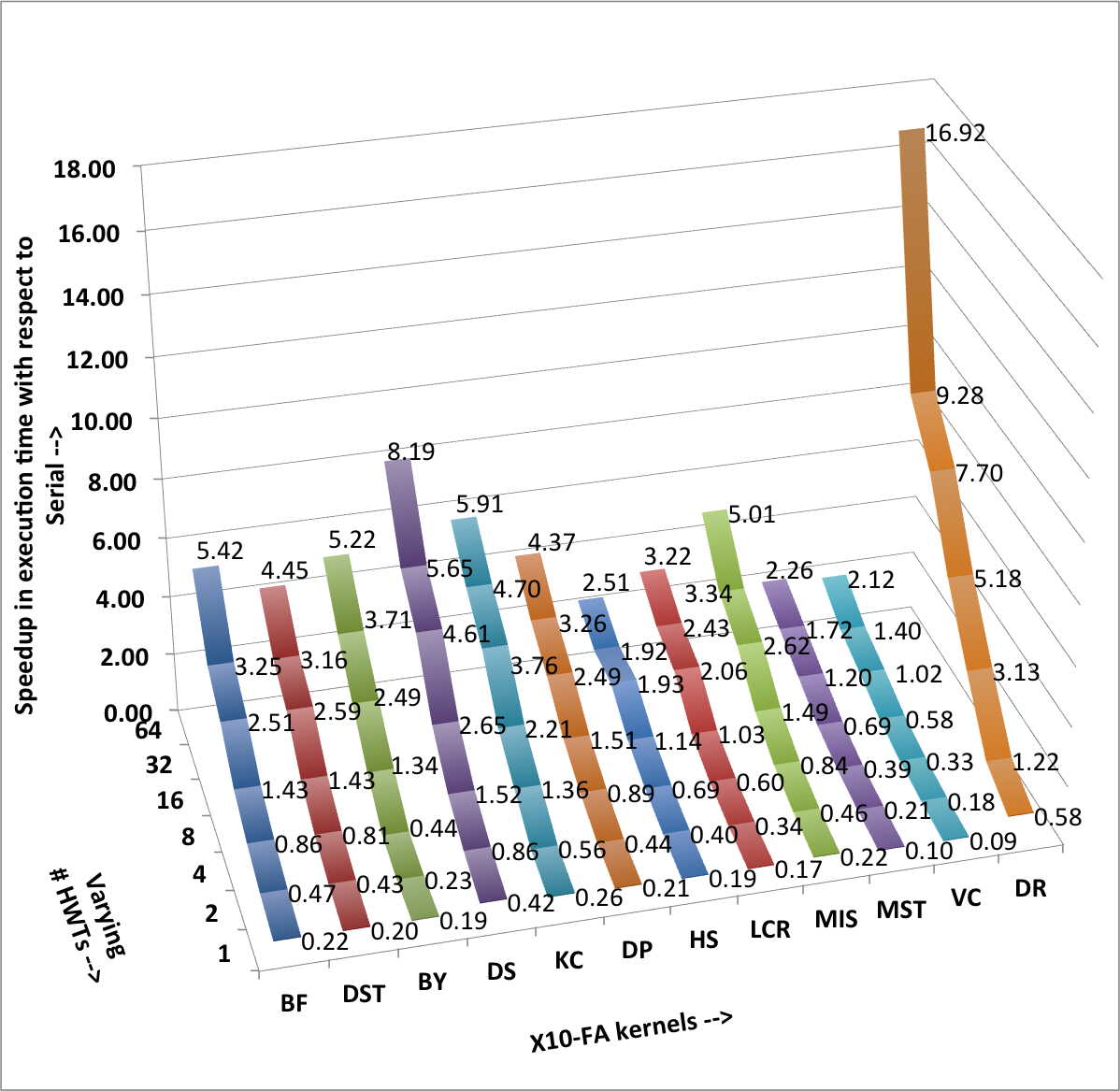}
                \caption{\fx10{} Max plot}
                \label{fig:X10max}
        \end{subfigure}%
        \begin{subfigure}{0.4\textwidth}
                \includegraphics[height=0.8\textwidth,width=\textwidth]{X10_64_min.png}
                \caption{\fx10{} Min plot}
                \label{fig:X10min}
        \end{subfigure}

        \begin{subfigure}{0.4\textwidth}
                \includegraphics[height=0.8\textwidth,width=\textwidth]{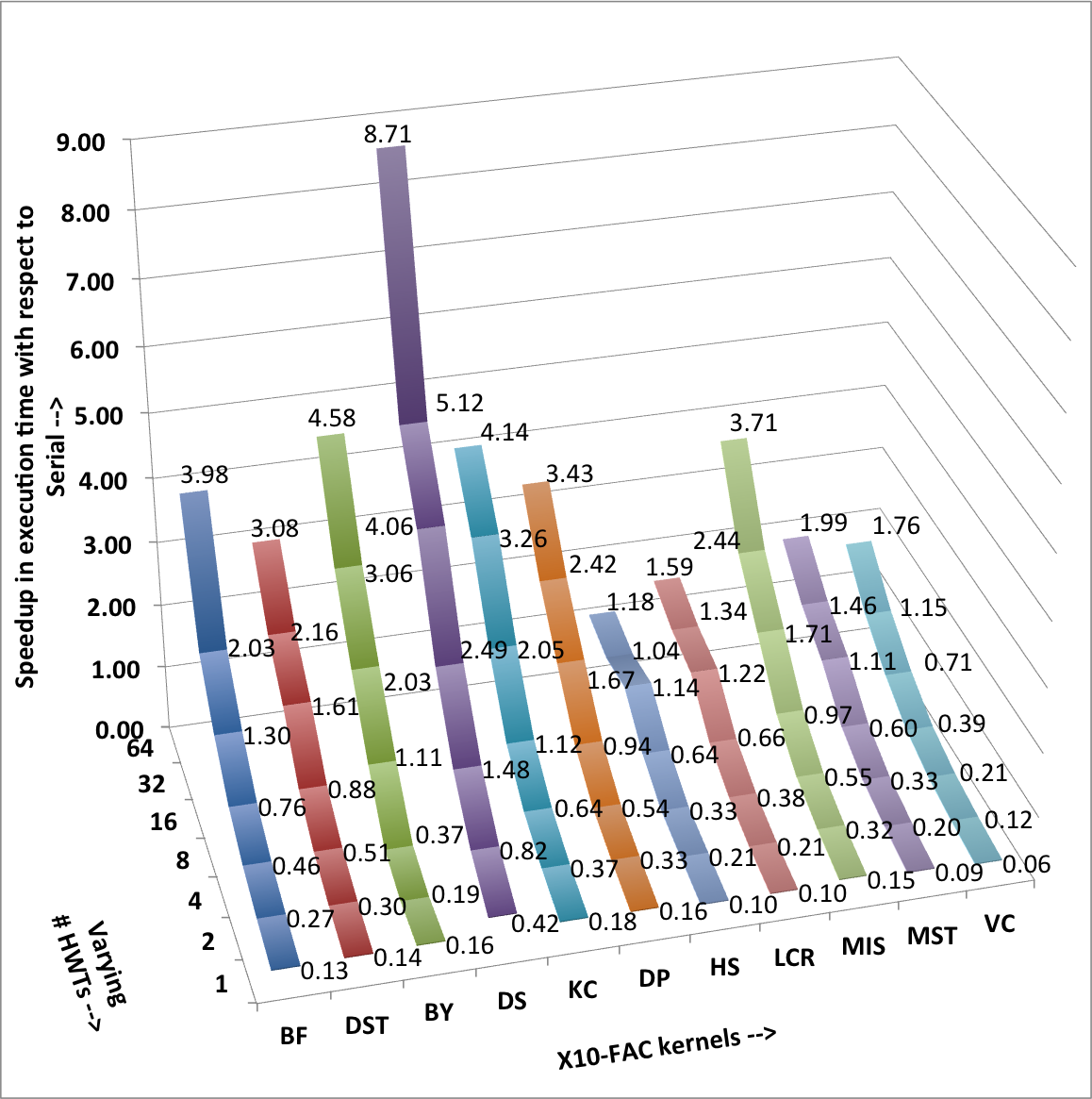}
                \caption{\cx10{} Max plot}
                \label{fig:Xclkmax}
        \end{subfigure}%
        \begin{subfigure}{0.4\textwidth}
                \includegraphics[height=0.8\textwidth,width=\textwidth]{XPhasers_64_min.png}
                \caption{\cx10{} Min plot}
                \label{fig:Xcklmin}
        \end{subfigure}
       
        \caption{{\fx10{} and \cx10{} plots for varying \#~HWTs}; input size = 64 nodes.}
        \label{fig:new-moores-law}

\end{figure*}
}

\subsubsection{Effect of varying the number of HWTs (input size
and number of clusters fixed)}
\label{sss:vh}
Figure~\ref{fig:X10max} and~\ref{fig:Xclkmax} present 
the execution time (speedup) statistics 
of the \fx10{} and \cx10{} kernels, for varying number of hardware threads (1 to 64) in multiples of two.
\ifConf{
For all these runs we set the input variation to \maxin{} and input size to 64 nodes.
We have also studied the behavior of these kernels for the \minin{} input variation  and have found it to be similar.
}
\ifTR{
The input size is set to 64 nodes for all these runs.
To study the impact of the input we make use of two types of inputs \minin{} and \maxin{}.
}
We plot the execution time of the kernels with respect to that of
the serial implementation in the \UP{} model.
These plots show that the overall performance for all the kernels improves with 
increase in the number of available HWTs.
However, for any specific kernel the quantum of improvement varies with the
chosen input and the number of available HWTs and their configuration.



The performance improvement is more or less linear when we increase the number
of hardware cores from 1 to 8 (intra-processor communication only) -- this
improvement is because of the increased sharing of workload among the HWTs.%
\ifConf{
\begin{figure}[H]
        \begin{subfigure}{\columnwidth}
\centering
                \includegraphics[width=0.7\textwidth,height=0.46\textwidth]{X10_64_max.png}
		\caption{\fx10{} plots; input size = 64 nodes, \#~clusters = 64.}
		\label{fig:X10max}
        \end{subfigure}%

        \begin{subfigure}{\columnwidth}
\centering
                \includegraphics[width=0.7\textwidth,height=0.46\textwidth]{XPhasers_64_max.png}
		\caption{\cx10{} plots; input size = 64 nodes, \#~clusters = 64.}
		\label{fig:Xclkmax}
        \end{subfigure}%

        \begin{subfigure}{\columnwidth}
\centering
                \includegraphics[width=0.7\textwidth,height=0.46\textwidth]{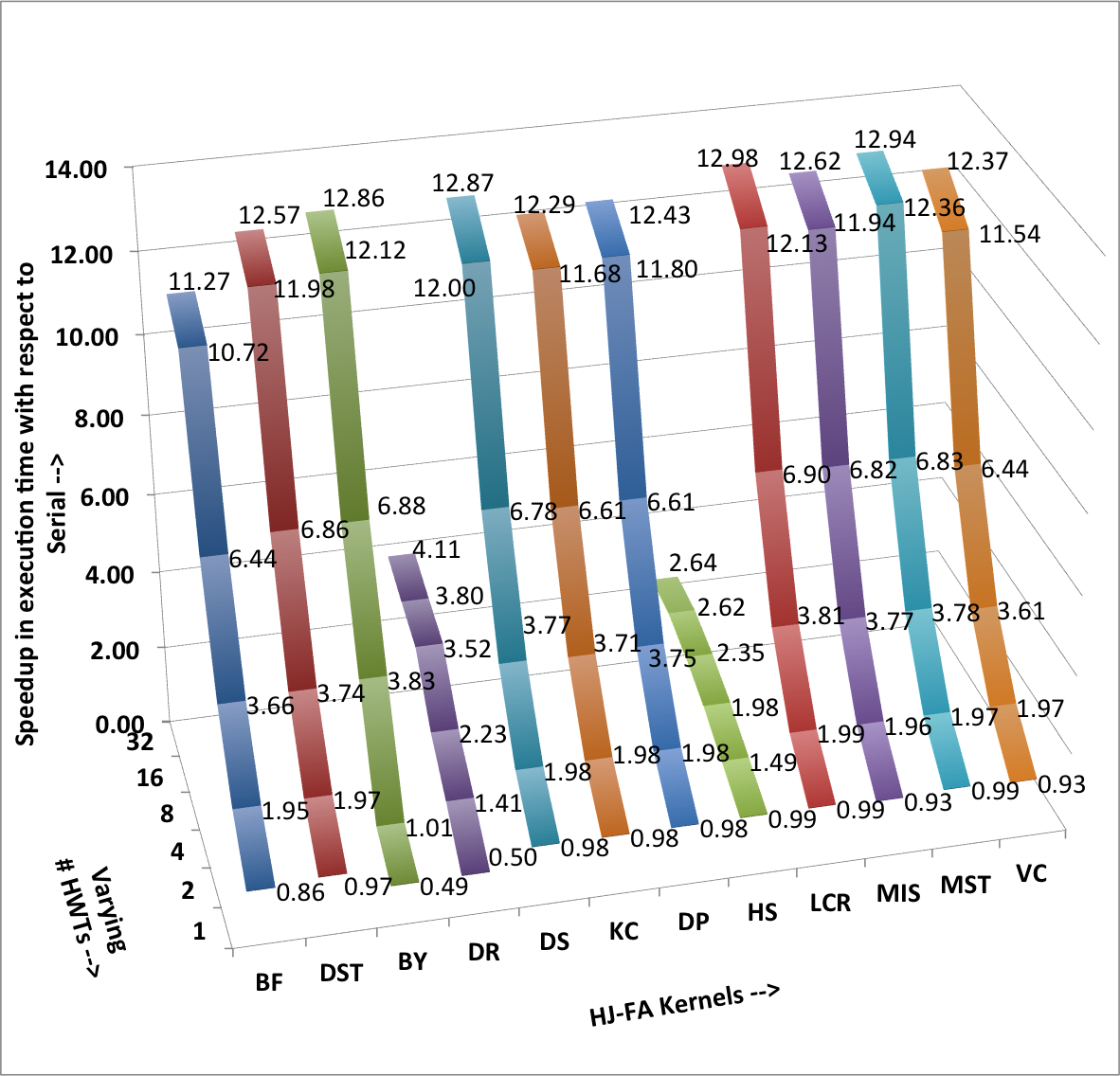}
        	\caption{\fhj{} plot; input size = 64 nodes, \#~clusters = 1, workload = 10 million instructions. }
        \label{fig:hjplots}
        \end{subfigure}%

	\caption{Effect of varying HWTs on \fx10{}, \cx10{}, and \fhj{} kernels.}
        \label{fig:new-moores-law}
\end{figure}
}%
On moving to 16 HWTs there is a slight dip in performance improvement (compared to 8
HWTs), owing to the inter-processor communication that comes into picture.
The performance improvement on 32 HWTs, compared to 16 HWTs is much less.
In case of 32 HWTs, while it doubles the sharing of workload by HWTs, 
it does not double other resources (for example, L1 cache).
As a result it incurs additional overheads due to increased conflicts in
accessing shared resources such as cache, interconnect, and so on.
This behavior is quite pronounced in {\em HS} where it leads to a slight
dip in performance (compared to 16 HWTs).
On going from 32 to 64 HWTs, the performance improvement depends on a host of factors --
increased communication cost (inter-hardware-node communication is more expensive than intra-node),
decreased scheduling overheads (each place runs on a unique HWT), 
decreased resource conflicts.
Depending on the specific kernel the effect varies. 
For example, in Figure~\ref{fig:Xclkmax},  {\em HS} shows  13\% improvement 
and {\em BF} shows 96\% improvement.


An interesting point to note is that 
in general for fewer hardware threads (1, 2, 4),
the serial versions in the \UP{} model runs faster than the
\fx10{} and \cx10{} versions.
This is due to the additional task creation, scheduling and termination overheads
present in these kernels.
As we increase the number of hardware threads (8, 16, 32, 64), the task
scheduling overheads decrease, and the effect of increased workload sharing
starts dominating the above mentioned overheads.

As discussed in Section~\ref{s:implement},
our kernels admit an additional option to introduce a user specified workload
in each asynchronous task.
We found that such an option is especially useful when our HJ based kernels are
simulated (on \SP{} model), where the time taken to execute these kernels is too
small (of the order few tens of milliseconds) to reason about the behavior of
these benchmarks; we tested the benchmarks for input size of 64 nodes.
To overcome this issue, we set a moderate workload of 10 million
instructions;
\ifConf{
Figure~\ref{fig:hjplots} shows a sample plot depicting the behavior the \fhj{} kernels,
for increasing HWTs, for the \maxin{} input.
}
\ifTR{
Figures~\ref{fig:hjplots} and~\ref{fig:phplots} show sample plots depicting the 
\fhj{} and \phj{} kernels for increasing HWTs, for the both \maxin{} and \minin{} inputs.
}
For brevity, we omit the plots of the \phj{} kernels as we found their
behavior to be similar.
Compared to the X10 based kernels, these HJ kernels admit
increased computational workload. 
This leads to a minor variation in their behavior compared to that of the
plots shown in Figure~\ref{fig:X10max}. 
For example, Figure~\ref{fig:X10max} shows a slight dip in 
in performance in {\em HS} in when we increase
the HWTs from 16 to 32; such a dip is not visible in Figure~\ref{fig:hjplots}.




\ifTR{
\begin{figure}[t]
        \begin{subfigure}{0.4\textwidth}
                \includegraphics[width=\textwidth]{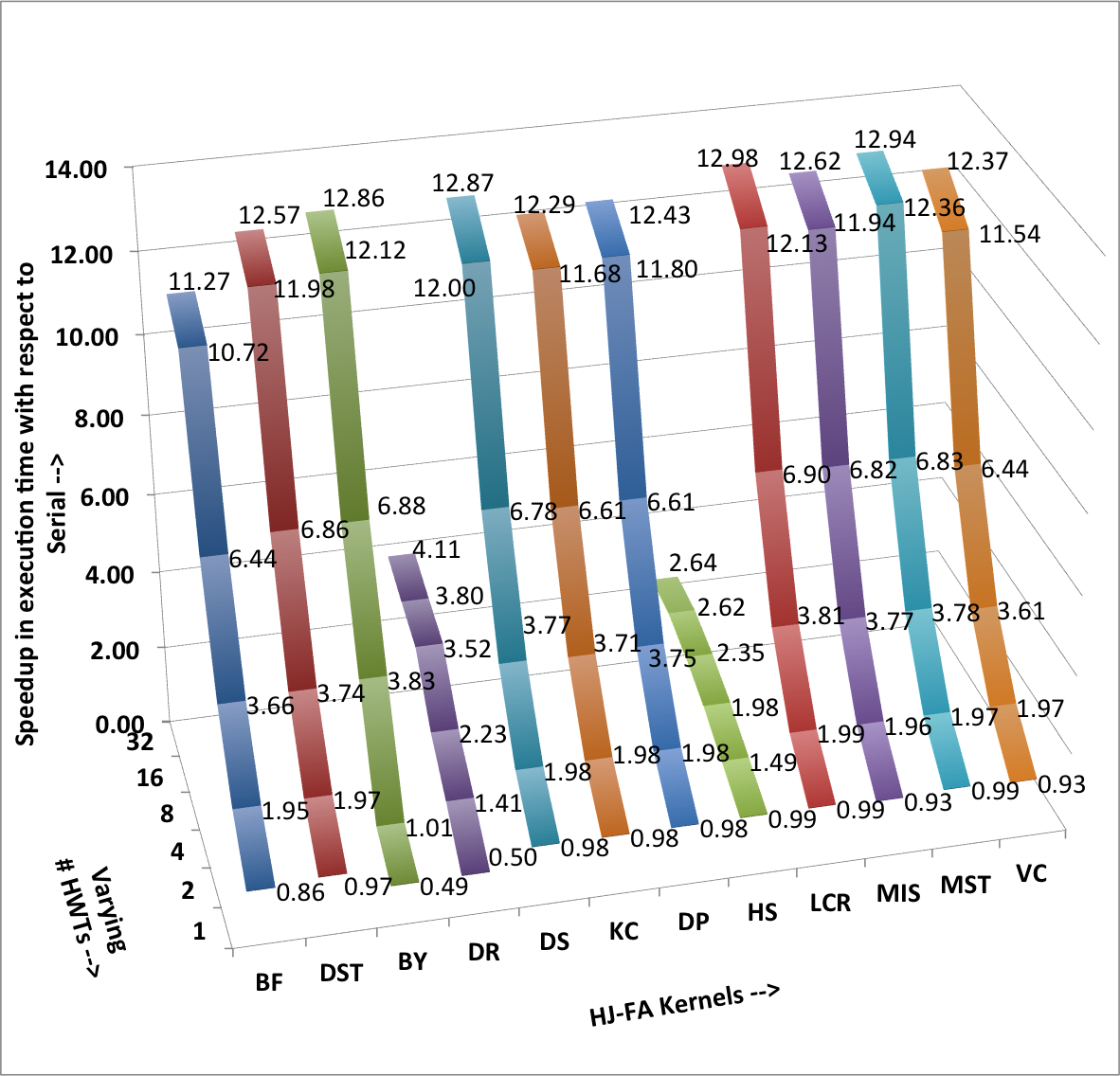}
                \caption{\fhj{} Max plot.}
                \label{fig:hjmax}
        \end{subfigure}%
        \quad
        \begin{subfigure}{0.4\textwidth}
                \includegraphics[width=\textwidth]{HJ_64_min.png}
                \caption{\fhj{} Min plot.}
                \label{fig:hjmin}
        \end{subfigure}
       
        \caption{{\fhj{} plots}; input size = 64 nodes, workload = 10 million instructions.}
        \label{fig:hjplots}
\end{figure}

\begin{figure}[t]
        \centering
        \begin{subfigure}{0.4\textwidth}
                \includegraphics[width=\textwidth]{Phaser_64_max.png}
                \caption{\phj{} Max plot.}
                \label{fig:phmax}
        \end{subfigure}%
        \quad
        \begin{subfigure}{0.4\textwidth}
                \includegraphics[width=\textwidth]{Phaser_64_min.png}
                \caption{\phj{} Min plot.}
                \label{fig:phmin}
        \end{subfigure}
       
        \caption{{\phj{} plots}; input size = 64 nodes, workload = 10 million instructions.}       
\label{fig:phplots}
\end{figure}
} 



\begin{figure}[t]
\begin{subfigure}{0.495\columnwidth}
\centering
                \includegraphics[width=\textwidth,height=0.7\textwidth]{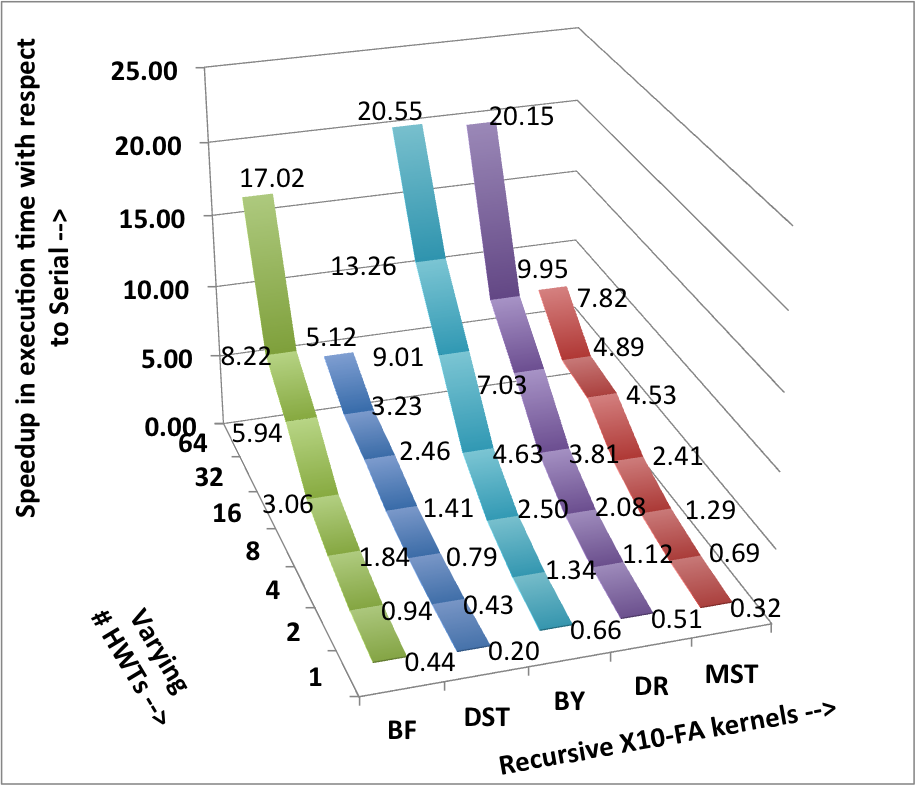}
                \caption{Recursive \fx10{} plot.}
		\label{fig:rec-x10-max}
\end{subfigure}
\begin{subfigure}{0.495\columnwidth}
\centering
                \includegraphics[width=\textwidth,height=0.7\textwidth]{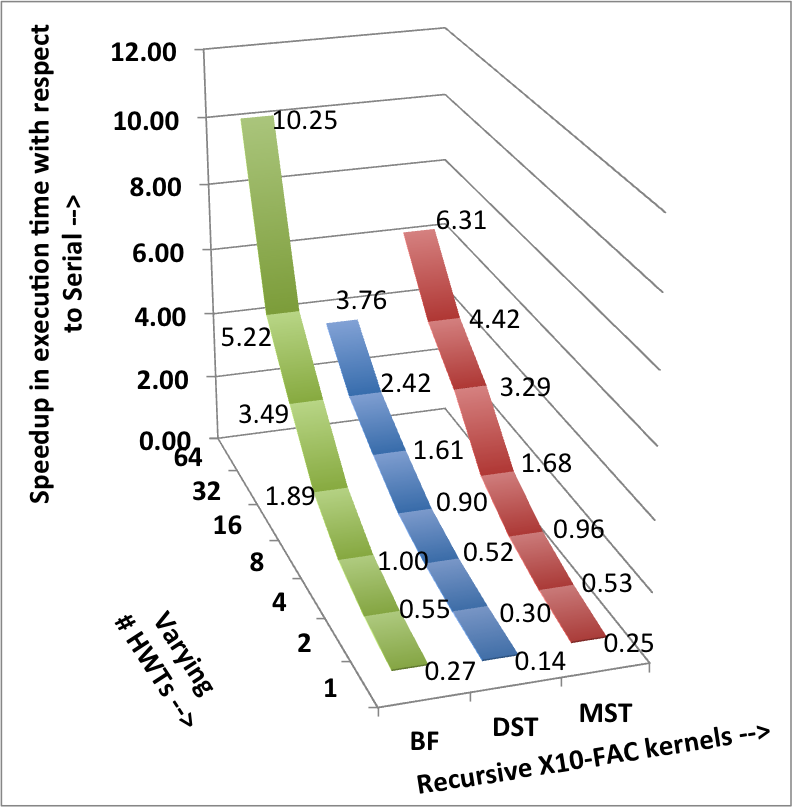}
                \caption{Recursive \cx10{} plot.}
		\label{fig:rec-cx10-max}
\end{subfigure}
\caption{\small Recursive kernels plots for varying \#~HWTs; input size = 64 nodes, \#~clusters = 64.}
\end{figure}

\begin{figure}
\begin{subfigure}{0.495\columnwidth}
\centering
                \includegraphics[width=\textwidth,height=0.7\textwidth]{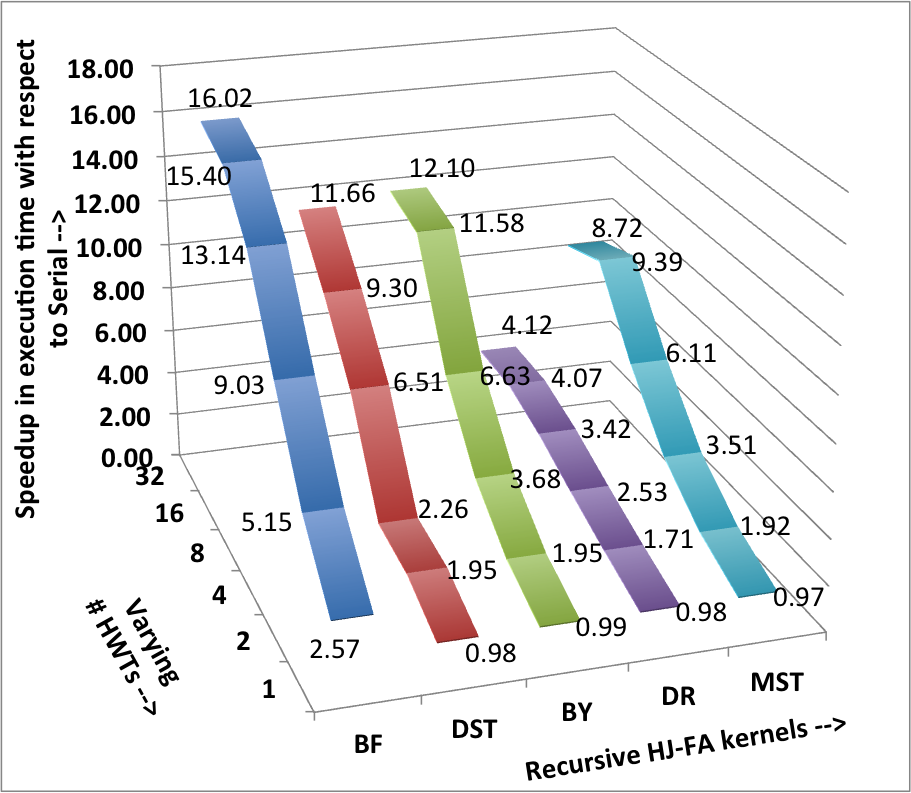}
                \caption{Recursive \fhj{} plot.}
		\label{fig:rec-hj-max}
\end{subfigure}
\begin{subfigure}{0.495\columnwidth}
\centering
                \includegraphics[width=\textwidth,height=0.7\textwidth]{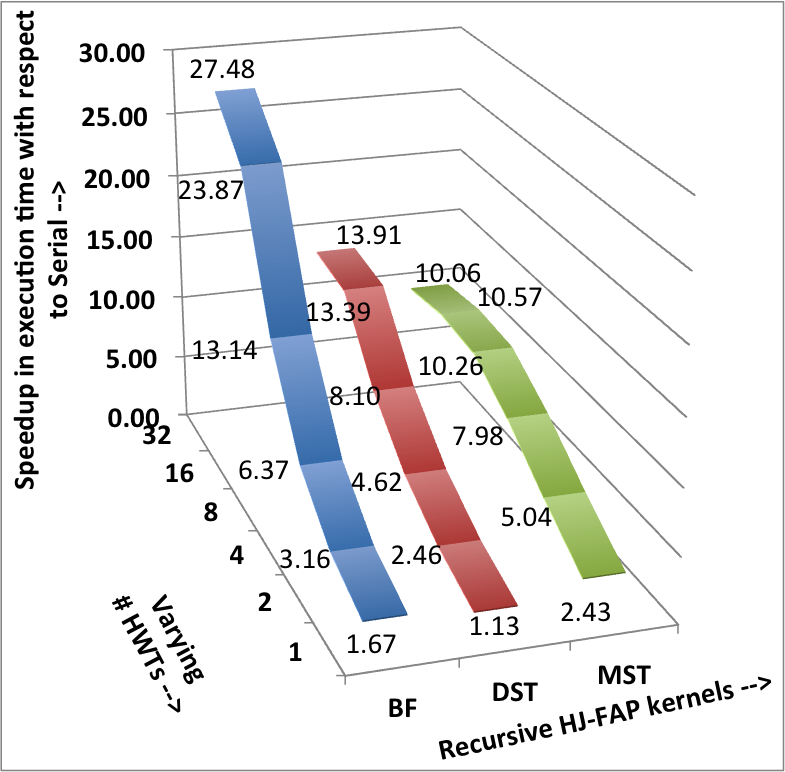}
                \caption{Recursive \phj{} plot.}
		\label{fig:rec-phj-max}
\end{subfigure}
\caption{\small Recursive kernels plots for varying \#~HWTs; input size = 64 nodes, \#~clusters = 64.}
\end{figure}

{\bf Recursive kernels:} Figures~\ref{fig:rec-x10-max} and~\ref{fig:rec-cx10-max} depict
the runtime characteristics of recursive \fx10{} and \cx10{} kernels, respectively,  with 
varying number of HWTs.
It can be seen that the behavior displayed by these kernels is similar to their
iterative counterparts.
Similarly Figures~\ref{fig:rec-hj-max} and~\ref{fig:rec-phj-max} depict the 
runtime characteristics of our recursive \fhj{} and 
\phj{} kernels, respectively with varying number of HWTs.
Similar to their iterative counterparts, we use a workload of 10 million 
instructions for these recursive HJ kernels.
It can be seen that performance for {\em MST} falls when the number of HWTs are 32 
as compared to 16 HWTs.
This is due to the relatively low value of workload in {\em MST}, which didn't offset the increased contention among HWTs (compared to the case where
the number of HWTs is set to 16) for the shared resources.
 
For brevity, in the rest of the section, we restrict ourselves to a subset of our
benchmark kernels.
We focus on the iterative \fhj{} and
\phj{} kernels when the number of clusters is set to 1
(Section~\ref{sss:vih}) and
on the iterative \fx10{} and \cx10{} kernels otherwise.


\begin{figure}[!t]
        \begin{subfigure}{\columnwidth}
\centering
                \includegraphics[height=0.46\textwidth,width=0.7\textwidth]{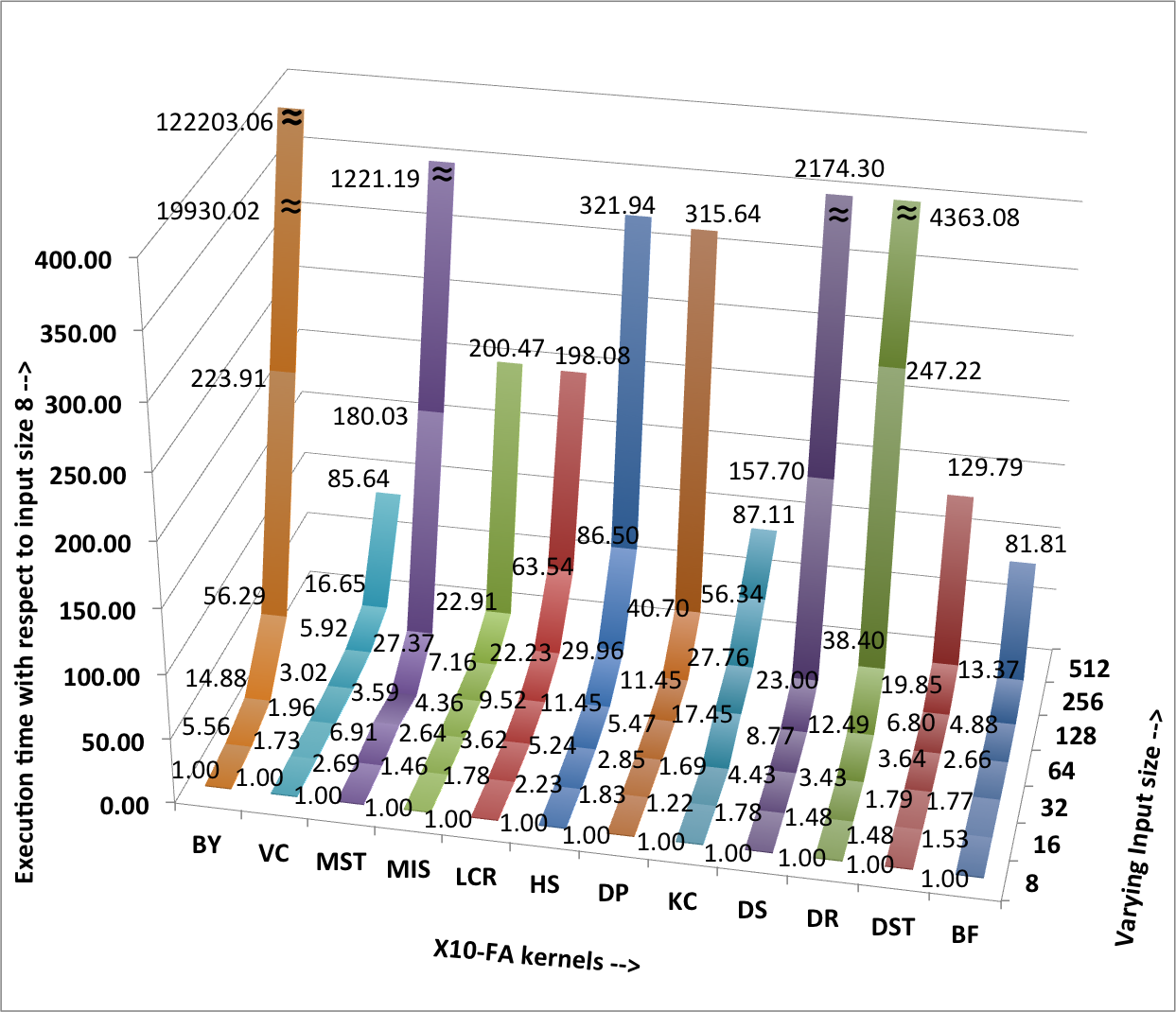}
                \caption{\fx10{}}
                \label{fig:ivX10}
        \end{subfigure}%

        \begin{subfigure}{\columnwidth}
\centering
                \includegraphics[height=0.46\textwidth,width=0.7\textwidth]{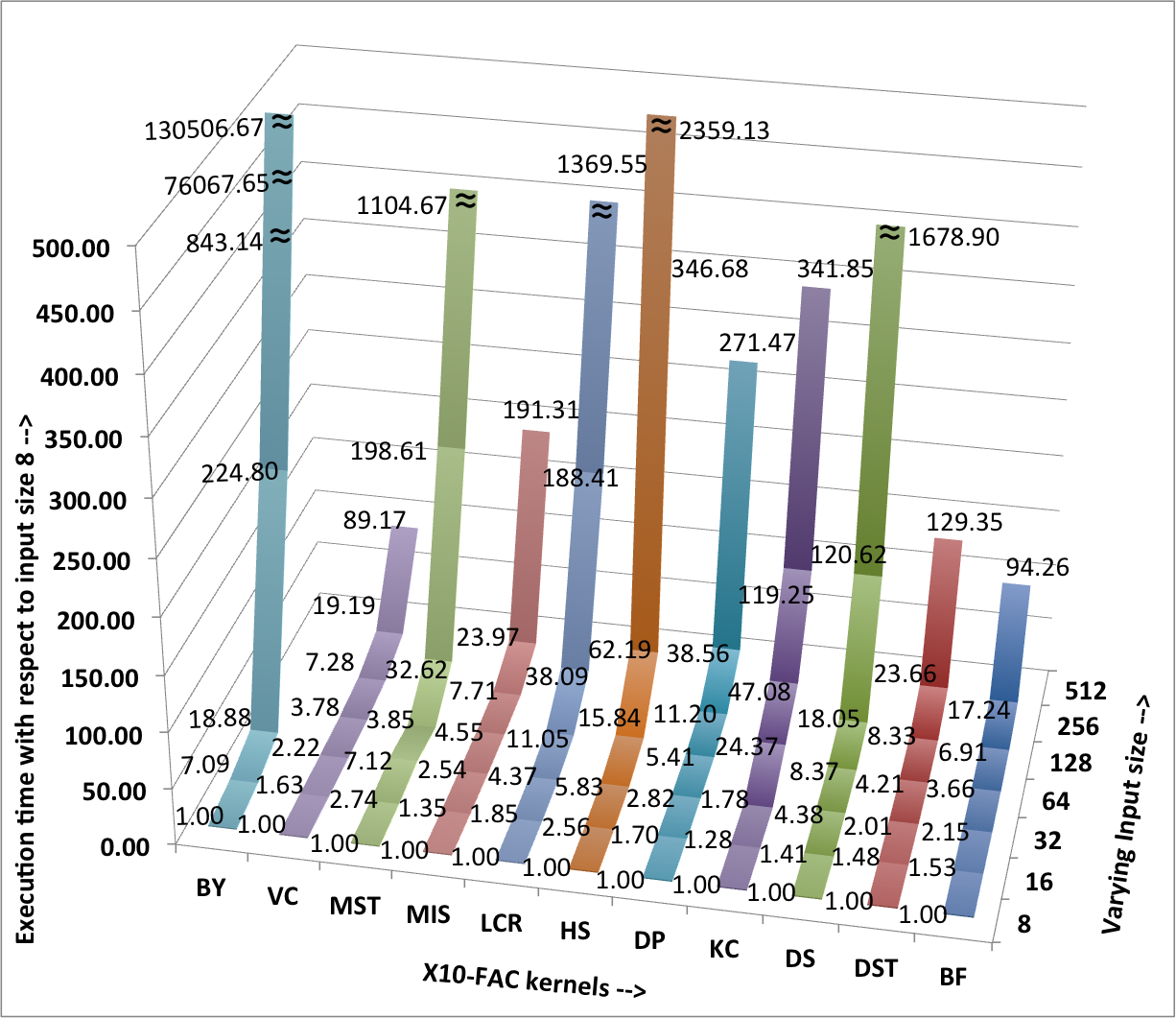}
                \caption{\cx10{}}
                \label{fig:ivXclk}
        \end{subfigure}%
        \caption{\small \fx10{} and \cx10{} plots for varying input size; \#~HWTs = \#~clusters = 8.
The execution times are normalized with respect to the execution time when
the input size is set to 8.
	}
        \label{fig:iv-x10}
\end{figure}

\subsubsection{Effect of varying the input size (number of HWTs and number
of clusters fixed)}
\label{sss:vi}
We vary the input size from 8 to 512 nodes, in multiples of two.
For our simulations we set both the number of clusters and HWTs to eight.
We choose eight HWTs for this study, as the communication between this  set of 
HWTs does not involve any inter-processor or inter-hardware-node communication.
For this study, the number of clusters could have been fixed at any one of
1, 2, 4, or 8 (if the number of clusters is more than eight then, for
smaller inputs, some clusters may not contain any nodes).
We broke the tie and set the number of clusters to 8; this leads to an
interesting configuration where each cluster of nodes is simulated on a unique
HWT and all the nodes in a cluster are simulated on a single HWT.
We study the effect of varying the number of clusters in Section~\ref{sss:vc}
and the combined effect of varying input size and number of clusters in 
Section~\ref{sss:vic}.

Figure~\ref{fig:iv-x10} presents the behavior of \fx10{} and \cx10{} 
kernels when run on the above-discussed configurations.
It is evident from the plots that as the input size increases, the 
execution time for the kernels also increases.
This behavior can be attributed to large inputs that lead to larger graphs
(hence higher memory requirement), 
increased phases/rounds and communication.
This effect is especially pronounced for {\em BY}, where the algorithm
requires a large number of rounds to execute and in each round every node
has to communicate with all the other nodes.


\subsubsection{Effect of varying the number of clusters (input size and
number of HWTs fixed)}
\label{sss:vc}
To study the effect of the clustering of nodes on the \benchname{}
kernels,
we vary the number of runtime places from 1 to 64.
We fix the HWTs to 8 (for the reasons mentioned above), and input
size to 64 nodes (to ensure that each cluster simulates at least one input node).
Figure~\ref{fig:pv-x10} presents the behavior of \fx10{} and \cx10{} 
kernels when run on the above-discussed configurations.


\begin{figure}[!t]
        \begin{subfigure}{\columnwidth}
		\centering
                \includegraphics[height=0.46\textwidth,width=0.7\textwidth]{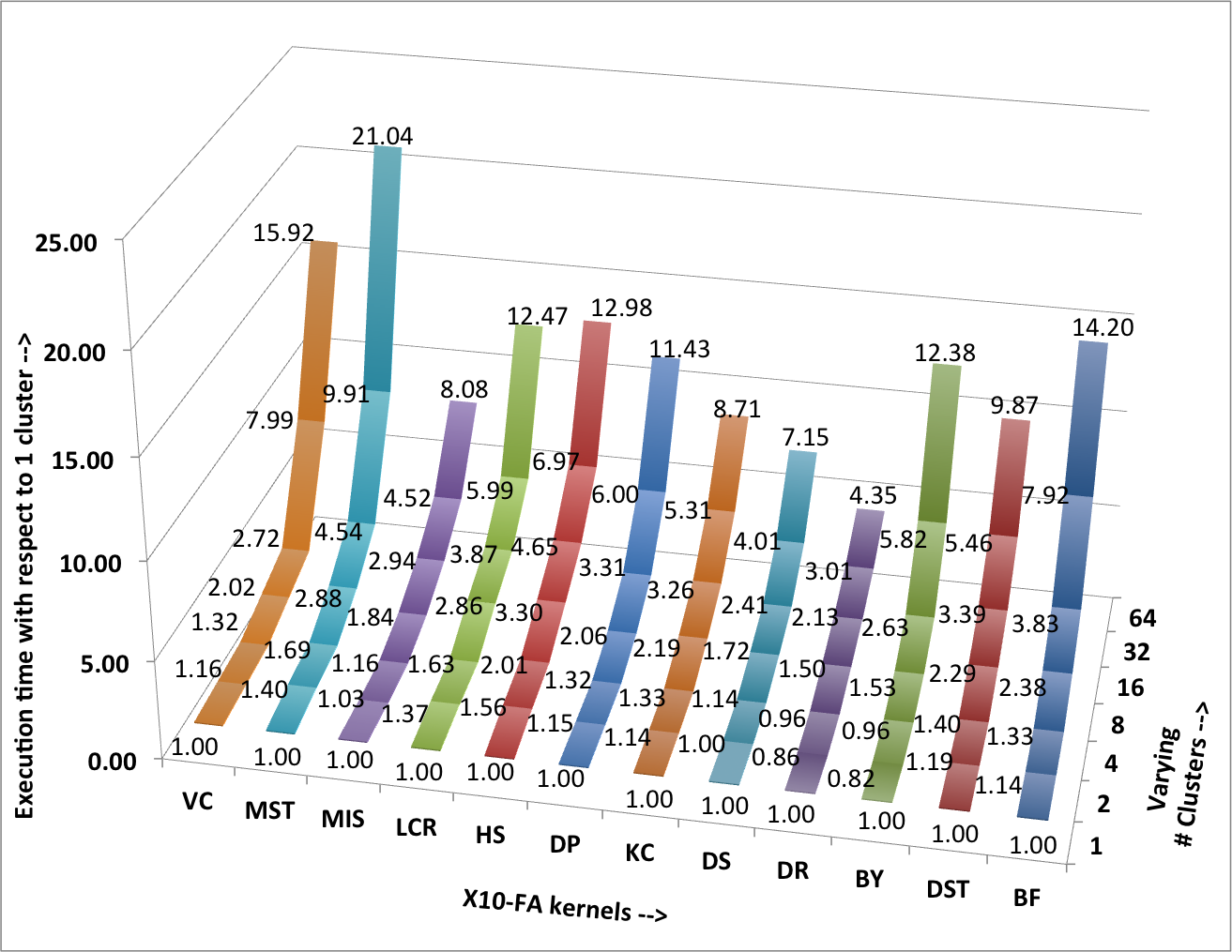}
		\caption{\fx10{}}
		\label{fig:pv-fx10}
	\end{subfigure}
        \begin{subfigure}{\columnwidth}
		\centering
                \includegraphics[height=0.46\textwidth,width=0.7\textwidth]{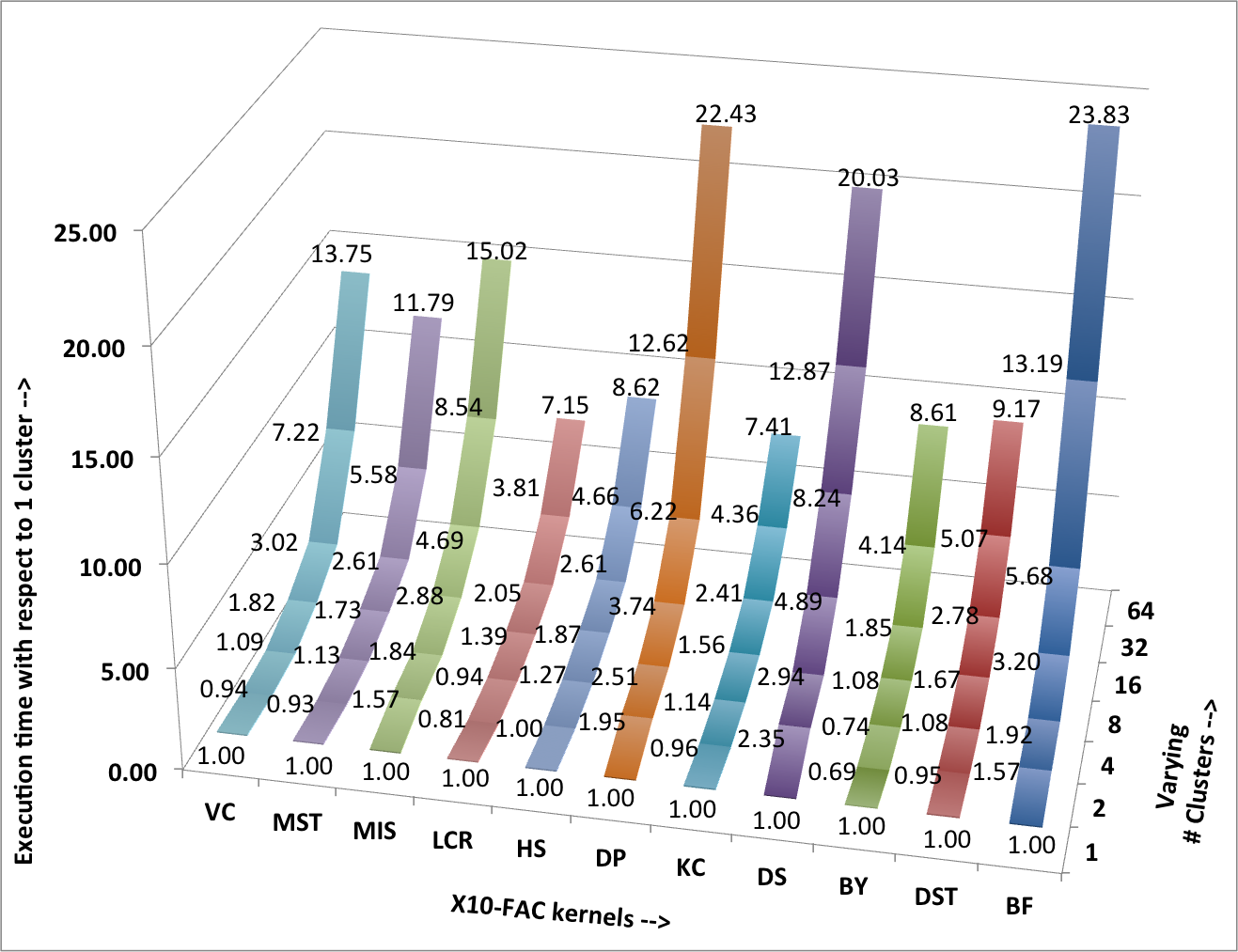}
		\caption{\cx10{}}
		\label{fig:pv-cx10}
	\end{subfigure}
        \caption{\small \fx10{} and \cx10{} plots for varying \#~clusters; input size = 64, \#~HWTs  = 8.
The execution times are normalized with respect to the time taken 
when the \# clusters is set to 1.  
	}
        \label{fig:pv-x10}
\end{figure}

We note that (for the most part) as the number of places increases
there is a proportional increase in the execution time of the benchmark
kernels; this is owing to the increase communication cost between the
tasks running on different places.
Note that {\em BY} and {\em DR} in Figure~\ref{fig:pv-fx10} and
{\em VC}, {\em MST}, {\em LCR}, {\em KC}, {\em BY} and {\em DST} in
Figure~\ref{fig:pv-cx10} show a
slight improvement in performance when the number of places is increased 
from 1 to 2.
Based on our discussions with X10 developers, 
we conjecture that such curious behavior could be attested 
to gains resulting from a
combination of inter-related factors concerning the distribution of 
tasks, such as change in cache access patterns and decreased contention in accessing the job queues
(more places $\Rightarrow$ more number of job queues for a given set of
jobs $\Rightarrow$ less contention for job selection).
When we further increase the number of places the increased inter-place
communication cost overshadows any such gains.


\begin{figure}[H]
        \begin{subfigure}{\columnwidth}
\centering
                \includegraphics[height=0.46\textwidth,width=0.7\textwidth]{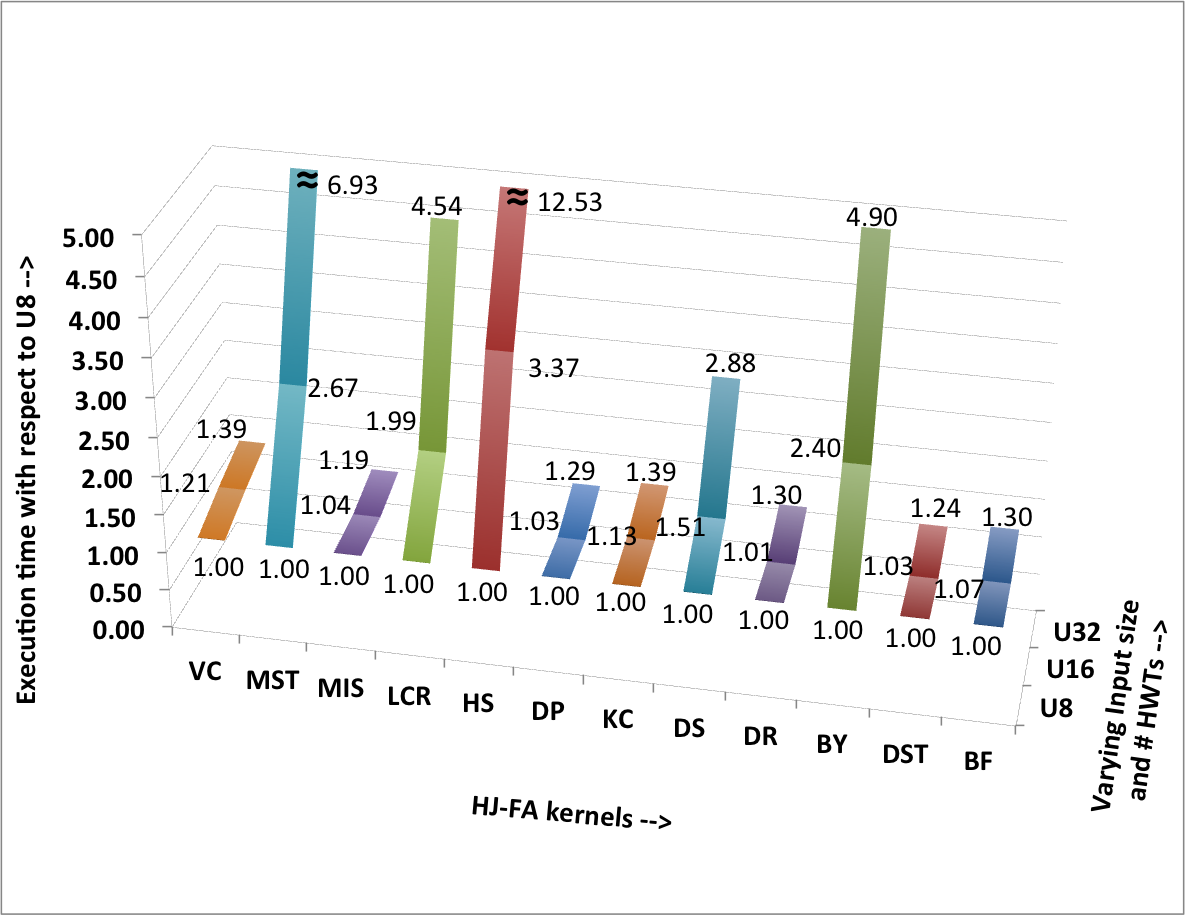}
                \caption{\fhj{}}
                \label{fig:vhjmax}
        \end{subfigure}%

        \begin{subfigure}{\columnwidth}
\centering
                \includegraphics[height=0.46\textwidth,width=0.7\textwidth]{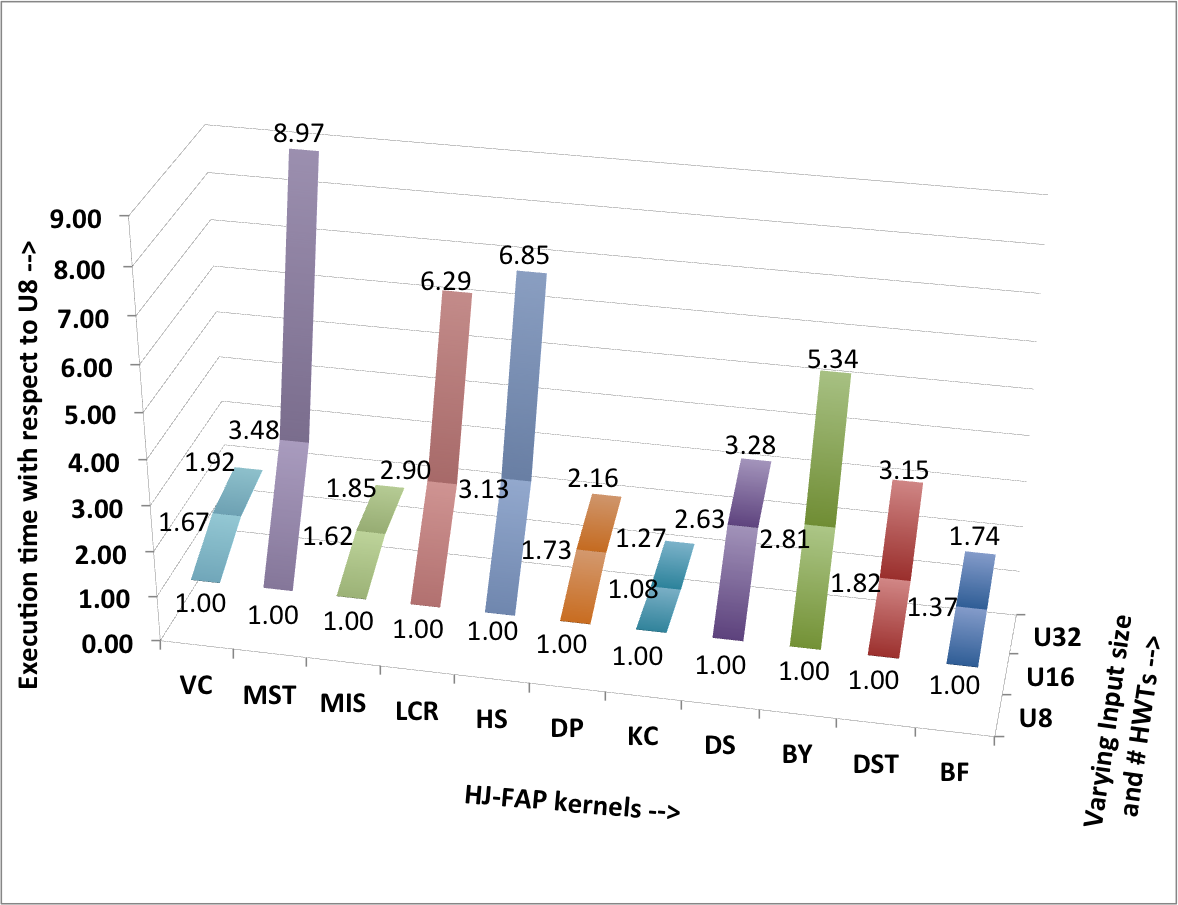}
                \caption{\phj{}}
                \label{fig:vphasermax}
        \end{subfigure}%
        \caption{\small \fhj{} and \phj{} plots for varying runtime configuration U$n$; $n$ = \#~HWTs  = input size; \#~clusters = 1.
The execution time numbers are normalized with respect to that of U$8$.
	}
        \label{fig:vhjplots}
\end{figure}

\subsubsection{Effect of varying the input size and number of HWTs (number of clusters fixed)}
\label{sss:vih}
We fix the number of clusters (places) to 1,
to avoid the costs incurred due to inter-place communication.
This setting also enables us to demonstrate the behavior of our HJ based 
kernels (\fhj{} and \phj{}); Figure~\ref{fig:vhjplots} shows the runtime 
characteristics of these kernels.
Considering the lower/upper limits on the input size and HWTs, 
we vary these key parameters between 8 to 32, in sync.
That is, when the input size is set to $k$, the number of HWTs is also
set to $k$; we represent this configuration as {U$k$}.


Note: one may naively assume that
increasing the input size (say from 16 nodes to 32 nodes) will not lead 
to an increase in the execution time provided there is a proportional 
increase in the number of HWTs (say from 16 to 32).
The behavior of our kernels show that such a hypothesis does not hold.
The execution time for all the kernel programs increases as we gradually 
move from U8 to U32.
This is because of the increased computation and communication at each
node, owing to the increased input.
Further, the rate of increase in execution time is less when we move from
U8 to U16, compared to the case when we move from U16 to U32.
This is because of the increased 
resource conflicts between the hardware threads, in the later case.
The exact quantum of increase depends on the working of the particular algorithm
(amount of communication, computation and so on).

\subsubsection{Effect of varying the input size and number of clusters
(number of HWTs fixed)}
\label{sss:vic}
Figure~\ref{fig:pi-x10} displays the characteristics of \fx10{} and 
\cx10{} kernels
when the number of HWTs is set to 8 and 
the input size and number of clusters are varied from 8 to 64 (in multiples of two) in sync.
That is, when the input size is set to $k$, the number of clusters is also
set to $k$; we represent this configuration as {V$k$}.

%
\begin{figure}[!b]
        \begin{subfigure}{\columnwidth}
\centering
                \includegraphics[height=0.442\textwidth,width=0.7\textwidth]{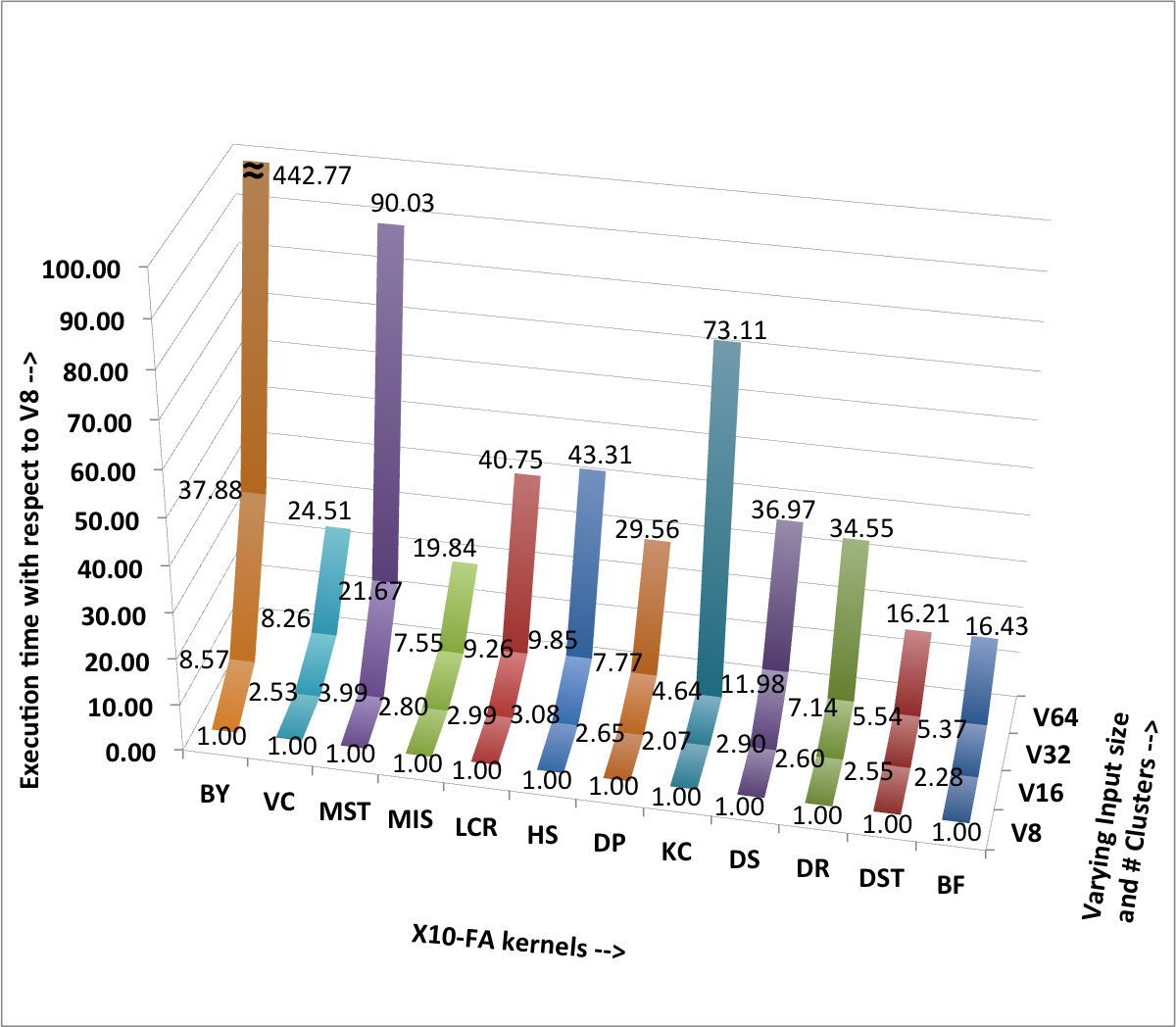}
                \caption{\fx10{}}
                \label{fig:pix10}
        \end{subfigure}%

        \begin{subfigure}{\columnwidth}
\centering
                \includegraphics[height=0.442\textwidth,width=0.7\textwidth]{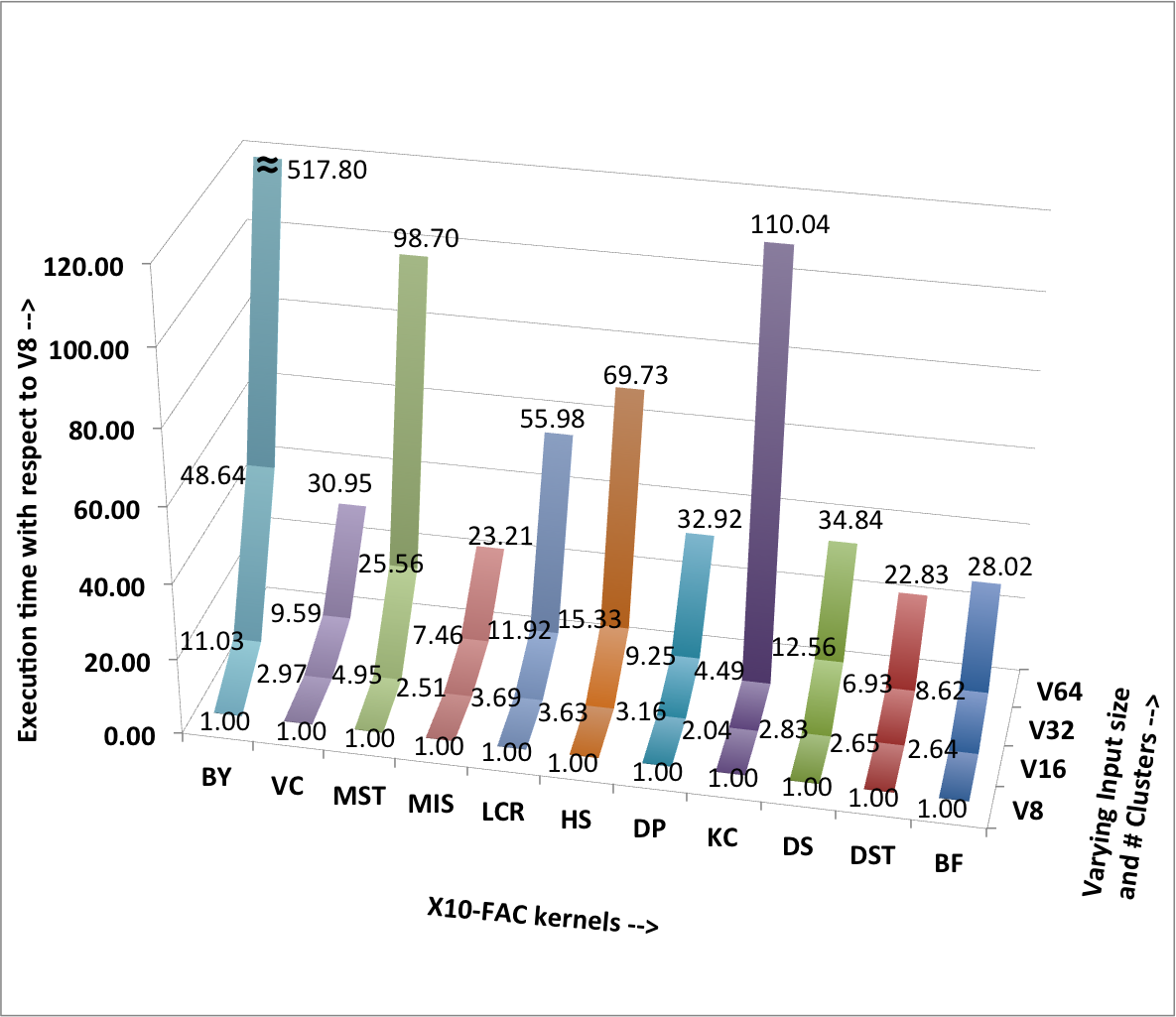}
                \caption{\cx10{}}
                \label{fig:pixclk}
        \end{subfigure}%
        \caption{\small \fx10{} and \cx10{} plots for varying runtime configuration V$n$; $n$ = \#~clusters = input size, \#HWTs = 8.
The execution time numbers are normalized with respect to that of V$8$.	}
        \label{fig:pi-x10}
\end{figure}

Note that the execution time of higher configurations (such as V64) is
significantly more than the lower ones (such as V8). 
This is because of three factors: a) increase in input size leads to increased
amount of work, b) increase in number of nodes leads to increase in overheads due
to conflicts in accessing shared resources (such as hardware threads,
memory and so on) by the tasks created,
c) increased number of clusters (places) leads to
increased (inter-place) communication cost.

\subsubsection{Effect of varying the number of clusters and number of HWTs
(input size fixed)}
\label{sss:vch}
This analysis helps understand how the clustering and increased
hardware threads affect the benchmarks. 
These characteristics have already been studied in Sections~\ref{sss:vh}, ~\ref{sss:vc} and ~\ref{sss:vih}.
We avoid further analysis of the same, for brevity.

\begin{figure}[!b]
        \begin{subfigure}{\columnwidth}
\centering
                \includegraphics[height=0.426\textwidth,width=0.7\textwidth]{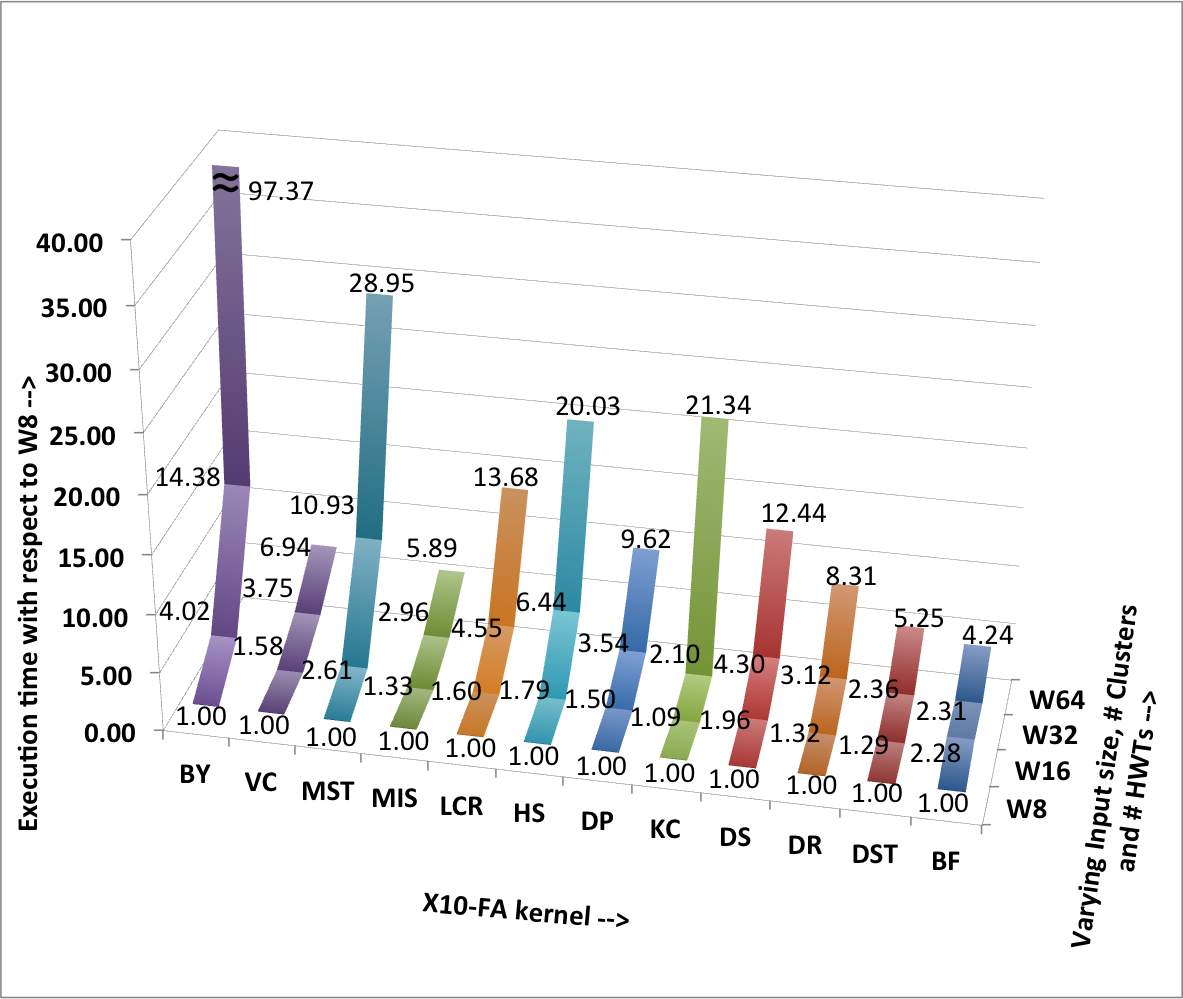}
                \caption{\fx10{}}
                \label{fig:vX10max}
        \end{subfigure}%

        \begin{subfigure}{\columnwidth}
\centering
                \includegraphics[height=0.426\textwidth,width=0.7\textwidth]{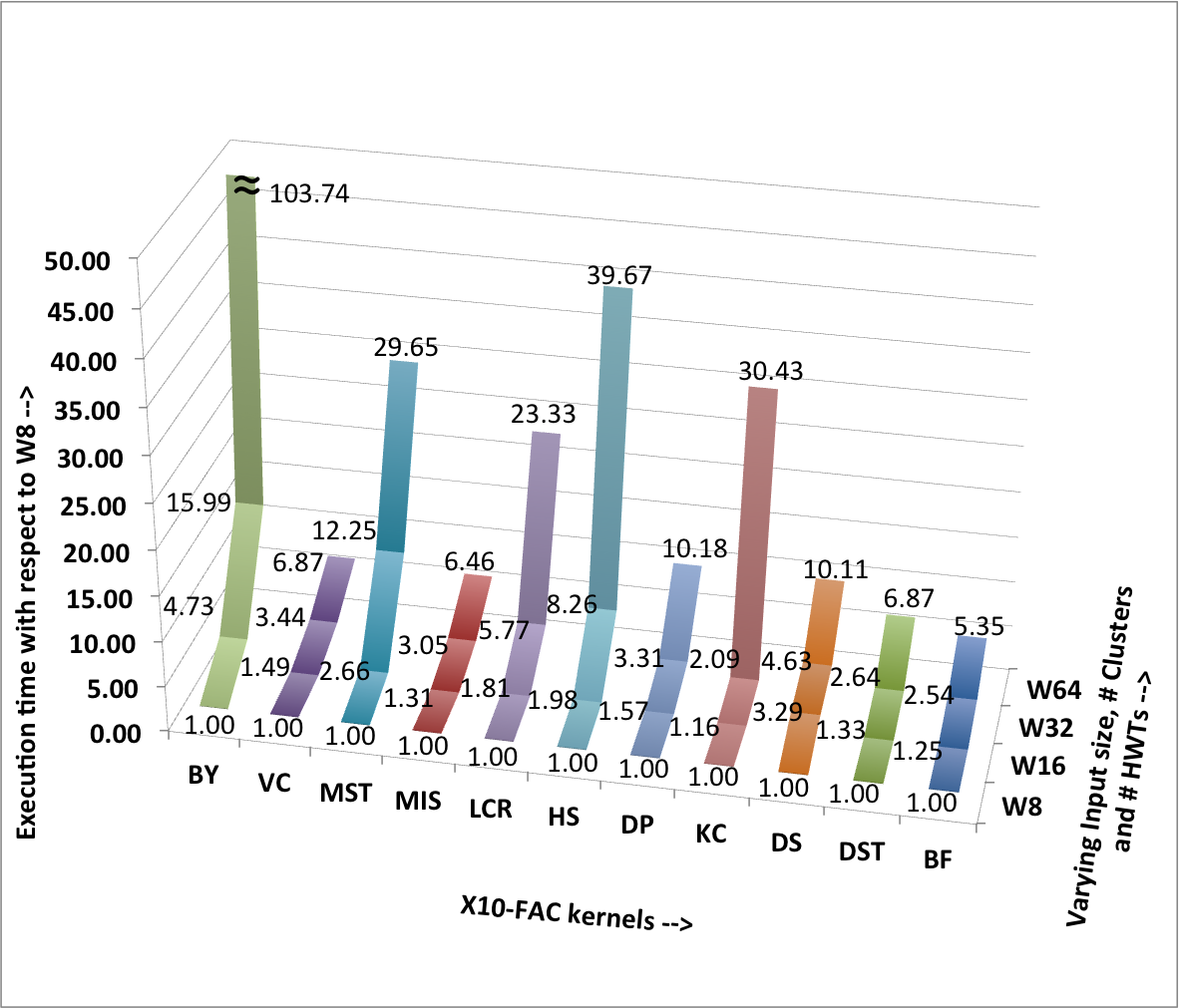}
                \caption{\cx10{}}
                \label{fig:vXclkmax}
        \end{subfigure}%
        \caption{\small \fx10{} and \cx10{} plots for varying runtime configuration W$n$; $n$ = \#~HWTs  = \#~clusters = input size.
The execution time numbers are normalized with respect to W8.
	}
        \label{fig:vx10plots}
\end{figure}

\subsubsection{Effect of varying the input size, number of clusters and number of HWTs}
\label{sss:ich}
Figure~\ref{fig:vx10plots} shows the effect of varying all the three key parameters in sync:
for running a kernel for input consisting of $k$ nodes, we consider each
node to be present in a unique cluster (thus, leading to $k$ clusters),
and the created tasks are run on $k$ HWTs; we represent this runtime configuration as {W$k$}.
\ifTR{
Figures~\ref{fig:vx10plots} and~\ref{fig:vx10clkplots} show these plots for both 
the \minin{} and \maxin{} input.
}
\ifConf{
}

In addition to the cost factors discussed in Section~\ref{sss:vih}, 
we now incur an additional cost (resulting from inter-hardware-node
communication) when we go from 32 to 64 HWTs.
These cost factors explain the increase in program execution time as we go
from W8 to W32 and a sharper increase as we go from W32 to W64.
The exact quantum of increase depends on the working of the particular algorithm
(amount of communication, computation and so on).

\ifTR{
\begin{figure*}[t]
        \begin{subfigure}{0.5\textwidth}
                \includegraphics[height=0.9\textwidth,width=\textwidth]{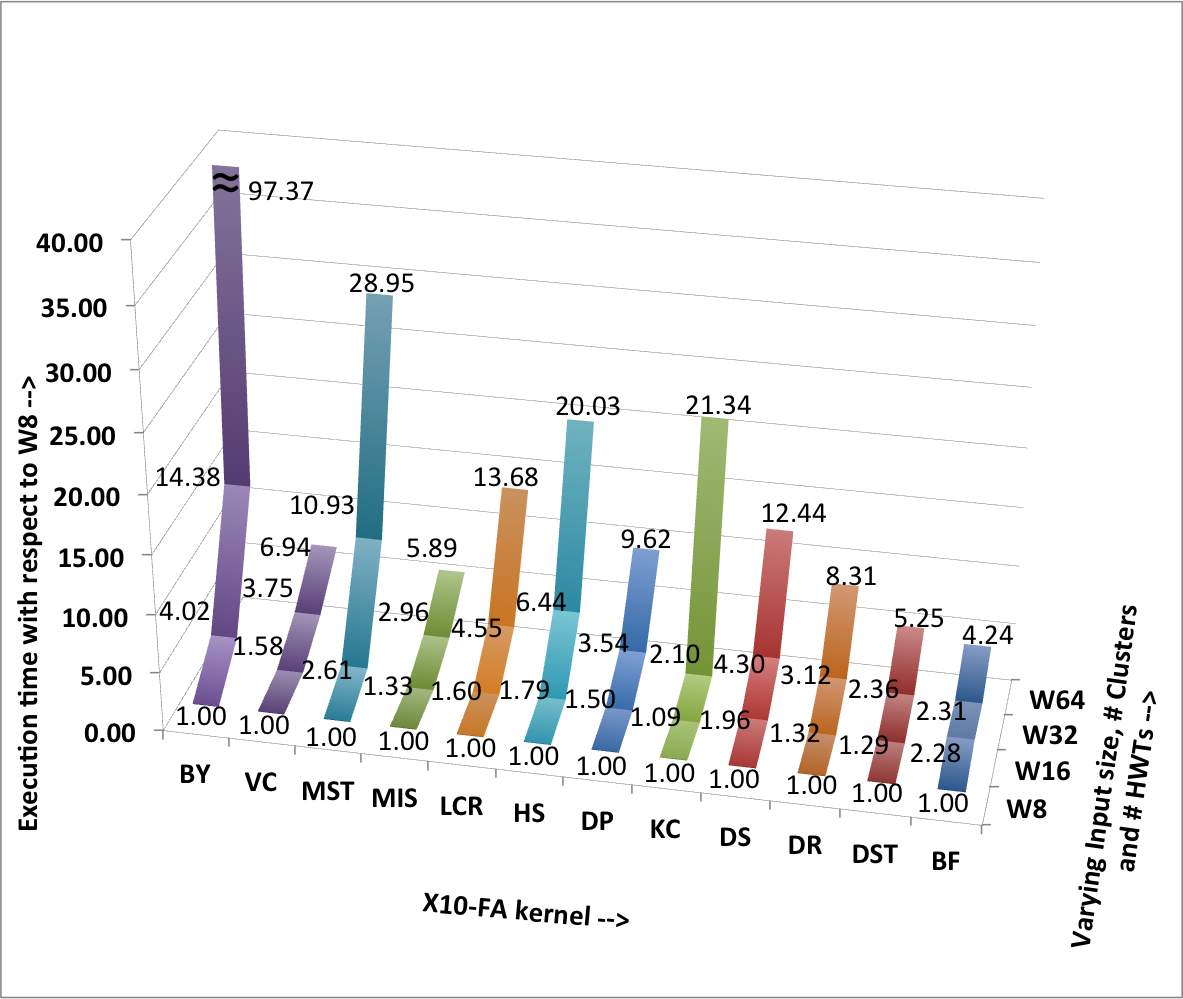}
                \caption{\fx10{} \maxin{} variation.}
                \label{fig:vX10max}
        \end{subfigure}%
        \begin{subfigure}{0.5\textwidth}
                \includegraphics[height=0.9\textwidth,width=\textwidth]{output_x10_min.png}
                \caption{\fx10{} \minin{} variation.}
                \label{fig:vX10min}
        \end{subfigure}
       
        \caption{\fx10 plots with number of cores  = number of places = number of input nodes; Vn refers to the value of {\em n} i.e. 8, 16, 32 \& 64}
        \label{fig:vx10plots}
\end{figure*}

\begin{figure*}[t]
        \begin{subfigure}{0.5\textwidth}
                \includegraphics[height=0.9\textwidth,width=\textwidth]{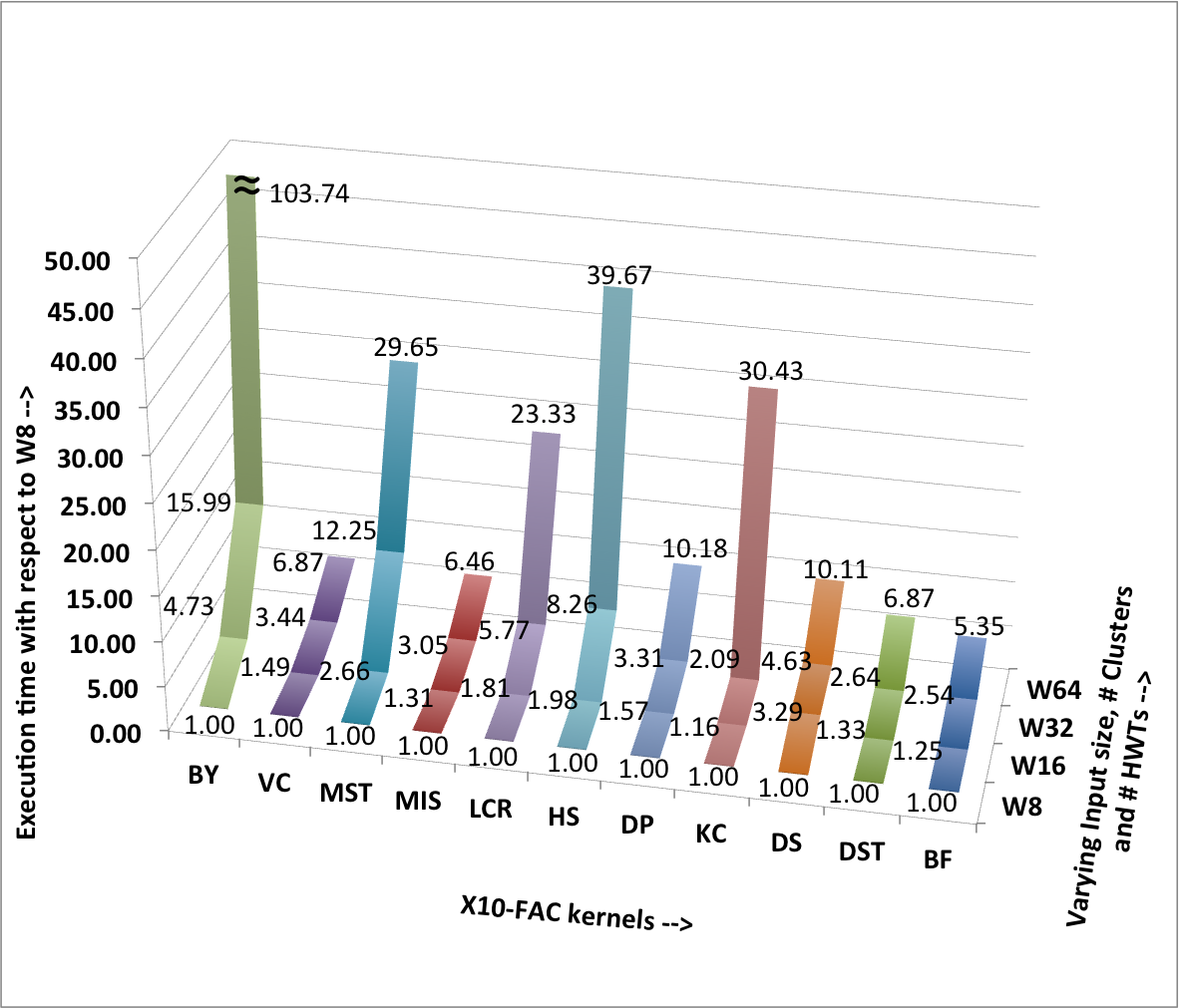}
                \caption{\cx10{} \maxin{} variation}
                \label{fig:vXclkmax}
        \end{subfigure}%
        \begin{subfigure}{0.5\textwidth}
                \includegraphics[height=0.9\textwidth,width=\textwidth]{output_xclocks_min.png}
                \caption{\cx10{} \minin{} variation}
                \label{fig:vXclkmin}
        \end{subfigure}
       
        \caption{\cx10 plots with number of cores  = number of places = number of input nodes; Vn refers to the value of {\em n} i.e. 8, 16, 32 \& 64}
        \label{fig:vx10clkplots}
\end{figure*}
}

\subsubsection{Effect of varying the input type}
\label{sss:vit}
For the \fx10{} and \fhj{} kernels,
Figure~\ref{fig:max-min} plots the ratio of the execution time of these kernels
for the input types \maxin{} and \minin{}.
We set the input size to 64 nodes and set the number of HWTs to 32
(the max number of HWTs that can be used by the HJ runtime on our hardware).
The number of clusters (runtime places) is set to 64 for \fx10{} kernels
and 1 for the \fhj{} kernels.

It can be seen that besides the time taken for performing the underlying computation,
the execution time of a benchmark is also dependent on (i) the number of task
creation and termination operations, and
(ii) the amount and cost of communication and mutex operations.
For example, as shown in  Figure~\ref{fig:fx10-characteristics}, in the 
{\em BY} kernel, the numbers of dynamic \finish{} and \async{} operations
for \minin{} input (longer diameter $D$) are more than that of \maxin{}
(shorter diameter $D$); this increases the execution time for the \fhj{} version. 
However, in the context of \fx10{}, the cost of decreased \finish{} and \async{} 
operations (in \maxin{}) gets shadowed by the increased cost of communication, 
which is significantly higher than \minin{}; this leads to a reversal of 
behavior.


\begin{figure}
\centering
\includegraphics[height=0.48\columnwidth,width=0.7\columnwidth]{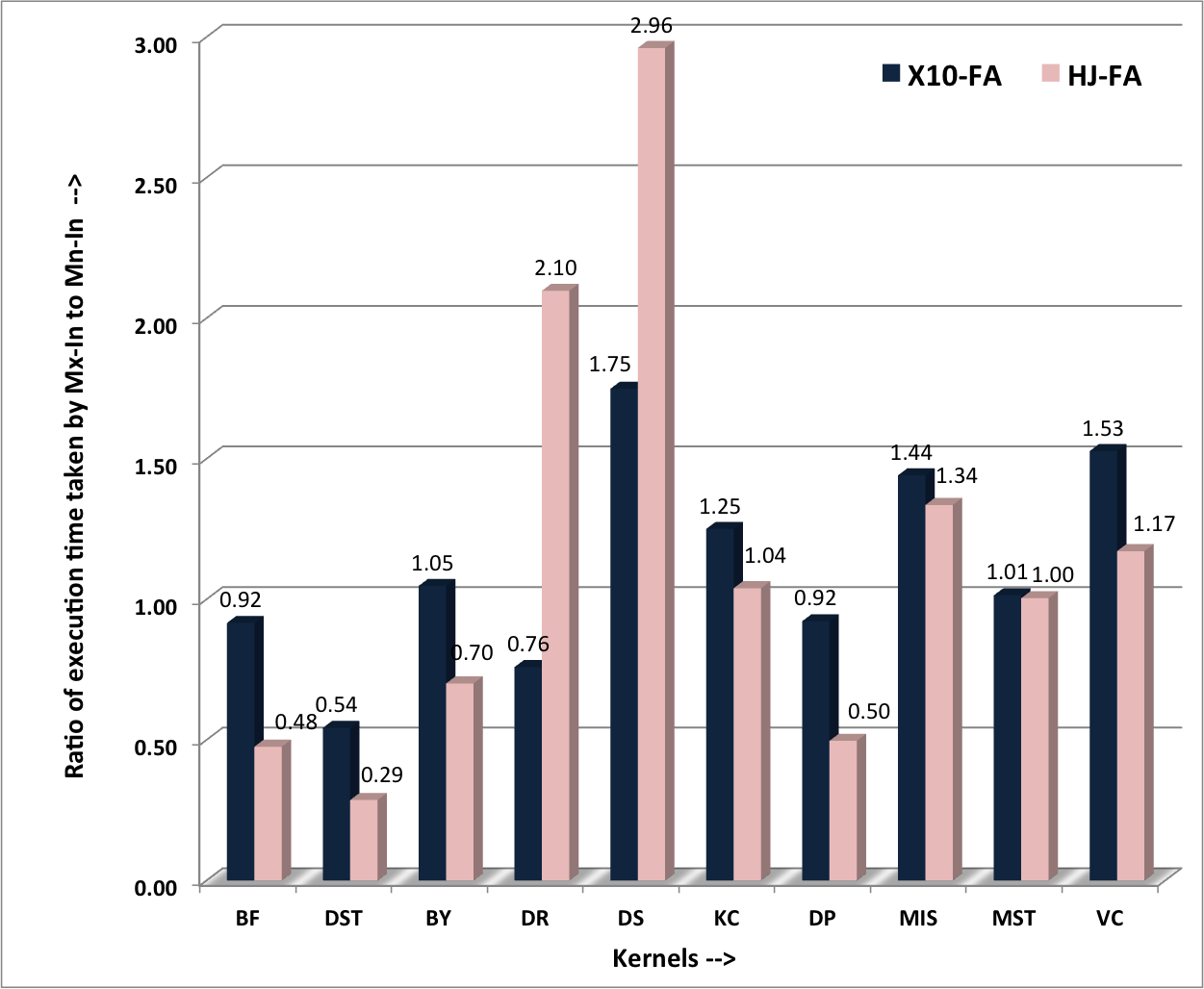}
\caption{\small \maxin{} Vs \minin{} for \fx10{} and \fhj{}; input size = 64, \#~HWTs = 32; \#~clusters = 1 (\fhj{}) and \#~clusters = 64 (\fx10{})}
\label{fig:max-min}
\end{figure}

\section{Scope of the Benchmarks}
\label{s:scope}
We now briefly discuss the scope of the \benchname{} kernels. 
We organize our discussion under four heads. 
\\
{\bf 1. Compiler optimizations and program analysis}:
An optimizing compiler can use the \benchname{} kernels to estimate its
effectiveness in optimizing distributed applications (involving remote
communication, barriers, load balancing and so on).
The metrics advocated by the \benchname{} kernels (such as number of task
creation, join, mutex, barrier operations) and our kernel behavior evaluation
schemes (distribution of communication; behavior with increase
in the number of hardware threads, number of clusters and input size) 
can give meaningful insights for designing new optimizations.
Further, the compiler writers can use these metrics and evaluation schemes to evaluate the overall effectiveness of their 
proposed optimizations.
Developers of new parallel and distributed program analysis techniques can use
the \benchname{} kernels as the basis to test the effectiveness (scalability,
precision and so on) of their proposed schemes.\\
{\bf 2. Runtime}: Developers of hypervisors, virtual machines and other managed
runtimes can use the \benchname{} to study and optimize the remote communications between 
different applications.\\
{\bf 3. Simulators}: 
Architecture and network simulators can use the communication trace generated
by the \benchname{} kernels, for varied inputs, to reason about the network traffic in the context of
varied distributed systems.\\
{\bf 4. Study of distributed systems}: 
Though our analysis is shown in the context of a tightly coupled system,
the \benchname{} kernels can be run on both tightly and loosely coupled systems.
This enables us to reason about different important aspects of distributed
systems, even in the absence of expensive distributed hardware.




\section{Conclusion}
\label{s:concl}
In this paper, we first identify a set of key requirements necessary for a kernel benchmark
suite implementing distributed algorithms. 
We then present and characterize a new kernel benchmark suite (named \benchname{}) 
that simulates twelve classical distributed algorithms (for varying input
and runtime configurations) and meets all the key requirements.
Currently, the kernels in \benchname{} are available in two task parallel
languages: X10 and HJ.
Considering the different popular parallel programming styles, we
present multiple variations of our kernels -- 31 parallel programs per
language.
To conveniently simulate varied configurations of distributed systems, we
present an input generator and an output validator for each algorithm under
consideration.
\benchname{} can be freely downloaded from
\url{http://www.cse.iitm.ac.in/~krishna/imsuite}.

Our proposed benchmarks can be ported to other languages that support 
distributed memory (such as MPI) and shared memory (such as OpenMP). 
We believe that our evaluation of \benchname{} in the context \UP{} and \SP{}
models of distribution, gives us some 
indication of how the future MPI and OpenMP ports of
\benchname{} may behave.

\newpage
\section*{\bf Acknowledgment}
We thank the anonymous reviewers for their insightful comments and
suggestions on an earlier version of this paper.
We thank T V Kalyan, Tripti Warrier and John Jose for their comments
on an earlier version of this paper.
We thank the X10 developers community, especially Igor Peshanksy and Vijay
Saraswat,  for their help in analysing some of the obtained results (chiefly
for the results in Section~\ref{sss:vc}).
This research is partially supported by the New Faculty Seed Grant, funded
by IIT Madras CSE/11-12/567/NFSC/NANV, DAE research grant CSE/13-14/139/BRNS/NANV and 
DST Fasttrack grant CSE/13-14/140/DSTX/NANV.

\bibliographystyle{elsarticle-num}
\bibliography{benchbib}

\end{document}